# High-strength and ductile lightweight cast aluminium alloys with superlattice nano-layered fibres (SNL) and core-shell nano-particles

Hemant Kumar[1*], Praveen Kumar[1], Dierk Raabe[2], Baptiste Gault[2,3*], Surendra Kumar Makineni[1*]

[1]Department of Materials Engineering, Indian Institute of Science, Bangalore 560012

[2] Max Planck Institute for Sustainable Materials, 40237 Düsseldorf, Germany

[3]Department of Materials, Royal School of Mines, Imperial College, Prince Consort Road, London SW7 2BP, United Kingdom

***Corresponding authors:**

hemantkumar2@iisc.ac.in, b.gault@mpie.de, skmakineni@iisc.ac.in

## Abstract

Lightweight, high-strength structural materials are component enablers in transportation and aerospace, reducing carbon footprints and enhancing fuel efficiency. Cast aluminium alloys, mainly based on eutectic compositions, make up ~85% of these materials but often fail catastrophically due to inefficient load transfer across the interfaces between the brittle eutectic phase and the ductile matrix. Here, we discovered that promoting a superlattice nano-layer (SNL) around the eutectic fibres, achieved by adding Zr to an Al-Gd near-eutectic alloy, enables excellent load transfer capabilities, resulting in a ~400% increase in tensile ductility. The primary α-Al matrix also contains a high number density of superlattice core-shell nano-particles. This exceptional increase in formability is attributed to the ability of the SNL to prevent dislocations from accumulating at the weak and brittle eutectic fibre/matrix interfaces, thereby avoiding stress concentrations that would otherwise initiate fibre breakage and debonding. The core-shell nano-particles in α-Al cause a large number of dislocation cross/multiple-slips on {111} planes, forming ultra-fine (~ 12 nm) dislocation networks that leverage substantial plastic strain accumulation. This atomic interface design overcomes the ductility limitations of cast-eutectic alloys, enabling them for structural applications.

***Keywords:*** *Lightweight Aluminium Alloys; Cast Eutectic Alloys; High-Temperature Strength; Ductility; Superlattice-Nano-Layer (SNL) Fibres; Core-shell Nano-particles; Dislocation Cells; High-Resolution Microscopy; Atom Probe Tomography.*

## Introduction

Aluminium (Al) alloys combine light weight with high specific strength, qualifying them as core structural materials in multiple strategic and engineering applications, ranging from safe aerospace parts to automobile powertrain components, poised to play an immense role in urgently needed emission reductions. Among these, cast Al-Si alloys with eutectic compositions are a significant fraction (~85%), mainly used for engine parts such as cylinder heads, valve filters, etc[1–3]. The eutectic alloys have an in-situ formed microstructure that develops during solidification[4], and comprises brittle intermetallic phase fibres/plates/lamellae inside the soft α-Al matrix. Recently, turbocharging and new direct fuel injection techniques[5] have increased thermal and mechanical loads on engine components, with temperatures locally exceeding 250°C. Existing Al-Si cast alloys cannot sustain these high temperatures and creep at high rates[6,7]. Hence, some heavy-duty diesel engines use heavier cast-iron heads and blocks. The most obvious strategy for improving fuel efficiency is to replace heavy parts with lightweight Aluminium cast alloys that are mechanically robust at high temperatures.

In the last two decades, Al-Ni[8] and Al-Ce[9,10] base cast eutectic alloys have been explored as promising candidates for high-temperature applications, particularly Ce-added alloys, due to the very low solid-state diffusivity of Ce (~4.6 x $10^{-19}$ $m^2/s$ at 400°C[11]) in the Al matrix. The microstructure consists of $Al_3Ni$ rods and $Al_{11}Ce_3$ plates embedded in an α-Al matrix, respectively. These are brittle intermetallics and share incoherent interfaces with the soft matrix. The near eutectic composition of Al-6wt.%Ni[12,13] and Al-12.5wt.%Ce[14] shows an excellent combination of microstructural stability up to 400°C and creep resistance up to 300°C. However, the alloys suffer from very low 0.2% yield strength (YS), (~90 MPa for Al-Ni and ~57 MPa for Al-Ce)[9,15]. An approach to induce coherent nanometric strengthening particles of $Al_3Sc$ into the soft α-Al matrix[16–18] of the eutectic alloys was quite successful in the increase of YS (~95-115 MPa for Al-Ni and ~139-166 MPa for Al-Ce alloys), but with reduced deformation capability, i.e. with very low values of ductility (~1 to 6 %)[3,19]. This strength-ductility trade-off is a major shortcoming for applications as structural materials, hindering their effective utilisation as heat-resistant alloys. Here, the brittle intermetallics are prone to easy fracture[20], and additionally, the incoherent, weak interfaces are unable to effectively transfer load upon the application of stress. These interfaces act as high-stress concentration sites and give rise to a high propensity for crack nucleation/propagation, leading to a loss of ductility and promoting brittle or early fracture[21,22].

In this work, we discovered that promoting superlattice nano-layered (SNL) eutectic fibres in the soft α-Al matrix prevents the formation of high-concentration sites upon loading, thereby increasing ductility by 400% from ~4% to ~20%. This is demonstrated in a near-eutectic $Al_{97.5}Gd_{2.5}$ (AG) microalloyed with Zr, as in $Al_{97.35}Gd_{2.5}Zr_{0.15}$ (AGZ). A fine distribution of coherent core-shell nano-particles also precipitates, strengthening the primary soft α-Al matrix, which enables a tensile strength of ~295 MPa at room temperature and retains up to 130 MPa at 250°C. A combined atomic-scale structural and compositional analysis was conducted to understand the microstructural origins and the deformation behaviour. Using this approach to explicitly address the root cause of the material's failure can channel the production of new lightweight, high-strength, heat-resistant, ductile eutectic alloys for structural applications.

**Results**

## Microstructural development of Al-Gd-Zr eutectic alloy with superlattice nano-layered (SNL) fibres

Fig. 1a shows a low-magnification back-scattered-electron (BSE) image of the cast Al-Gd-Zr alloy microstructure, revealing the distribution of the eutectic structure (α-Al/$Al_3Gd$ fibres) and primary α-Al matrix. The area fraction of the eutectic structure is estimated to be ~ 55%. A similar eutectic microstructure in terms of their fraction, scale, and morphology was also observed in the binary Al-Gd alloy, as shown in Supplementary Fig. S1. The fibres in the eutectic structure usually crystallise with incoherent interfaces with α-Al matrix, but in a few instances, semi-coherent interfaces were also found. Fig. 1b-c shows a high-angle-annular-darkfield (HAADF) scanning transmission electron microscopy (STEM) image centred on an interface taken along $[1\bar{1}0]_{\alpha\text{-}Al}$ zone axis. The masking of 220 planes across the interface shows the presence of dislocations (red oval-shaped overlays) at nearly equal intervals (~5.1 nm). The atomic structure of the fibres was found to be hexagonal-closed-packed (hcp) $DO_{19}$ ordered (with the lattice parameters a = 0.6125 nm and c = 0.9944 nm with a stoichiometry of $Al_3Gd$.

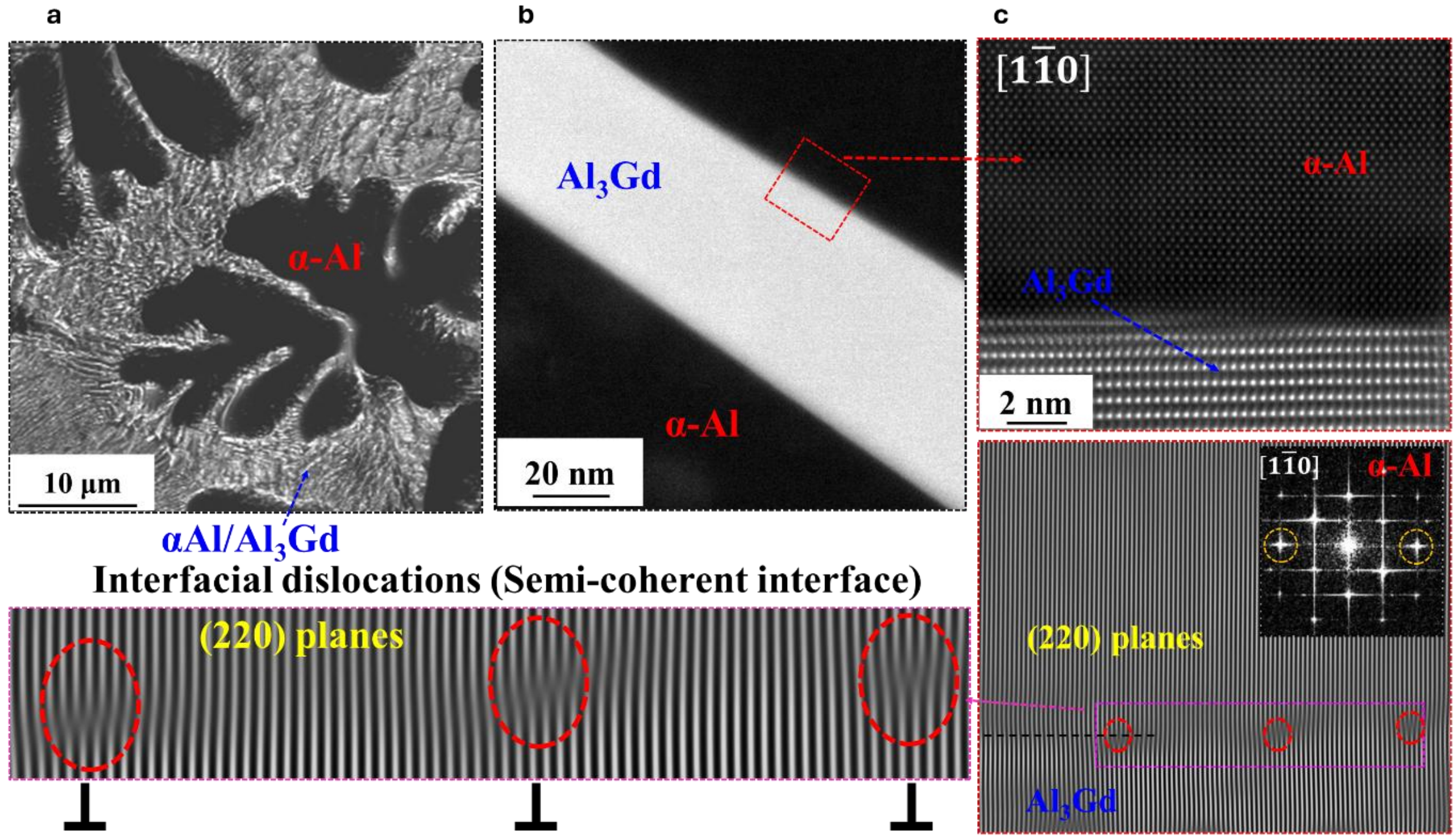

**Fig.1: Microstructure of cast Al-Gd-Zr (AGZ) alloy: a** A back-scattered-electron (BSE) scanning-electron-microscopy (SEM) image of the AGZ alloy showing the distribution of α-Al/$Al_3Gd$ eutectic regions and primary α-Al phase (darker contrast). **b** A low magnification and **c** a high-resolution HAADF STEM image taken across a fibre/α-Al interface in AGZ (without annealing) along $[1\bar{1}0]$ cubic zone axis. Masking of 220 planes from the Fast Fourier Transformation (FFT) along $[1\bar{1}0]$ zone axis (see inset) across the interface shows the presence of dislocations, indicating the semi-coherent nature of the interface.

The Al-Gd-Zr cast alloys were subjected to annealing at 400°C to promote the formation of a superlattice nano-layer envelope around the $Al_3Gd$ fibres. Fig. 2a shows the atom probe tomography (APT) reconstruction centred on an interface for the alloy with the corresponding Zr composition profile (Al and Gd profiles are shown in Supplementary Fig. S2) across the interface after 5 hours of annealing. The composition of the Gd-rich fibres is measured to be Al-25at.% Gd with a minute partitioning up to ~ 0.5at.% Zr with respect to the α-Al matrix. At the fibre/matrix interface, the Zr profile reveals segregation of up to ~2.5 at.% Zr.

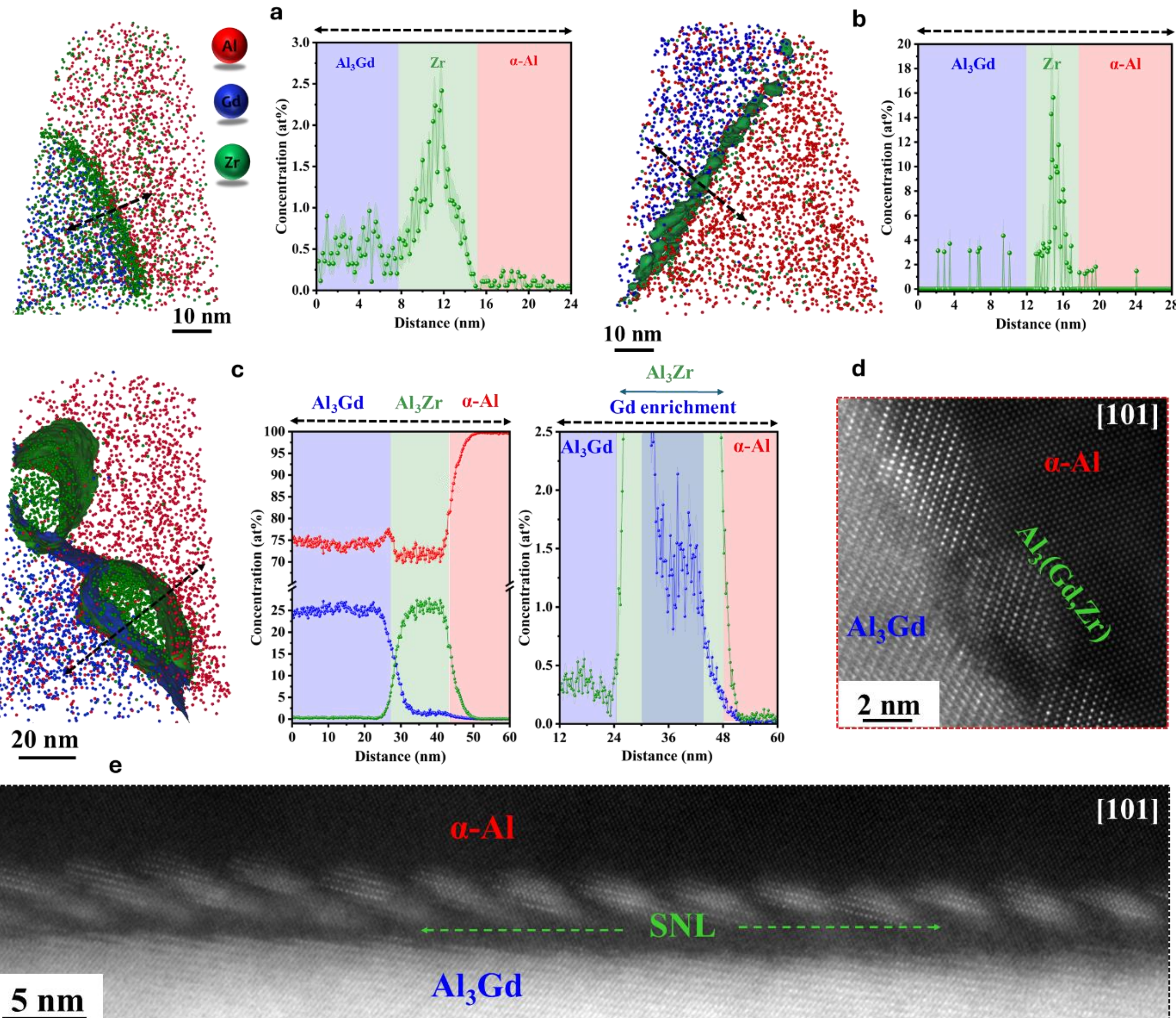

**Fig.2: Formation of Superlattice Nano-Layer (SNL) enveloping the $Al_3Gd$ eutectic fibres in AGZ alloy:** Atomic-scale compositional profiles across fibre/α-Al interface in Al-Gd-Zr alloy after annealing at 400°C for **a** 5 hours, showing segregation of Zr up to 2.5 at.% and **b** 10 hours**,** showing an increase in Zr up to 16at.% at the interface, indicating the early stages of formation of the SNL. **c** After 25 hours, Zr composition increases to ~ 25at.% with Gd partitioning of ~ 1.5at.%. **d** an atomic-scale high-angle-annular-darkfield (HAADF) STEM image centred on the fibre/α-Al interface for 25 hours annealed alloy showing the presence of $Al_3$(Zr, Gd) SNL having the fcc-based $L1_2$ crystal structure. **e** A lower magnification HAADF STEM image showing a continuous envelope of SNL at the interface.

The segregation is induced by annealing at 400°C, as confirmed by compositional analysis across the fibre/matrix interface in the AGZ alloy (without annealing), which reveals no Zr segregation at the interface (Supplementary Fig. S3). As the annealing time proceeds, the Zr composition increases locally along the interface, as shown in Fig. 2b (up to ~16 at.% after 10 hours), forming semicylinder-like Zr-rich regions (Fig. 2c) with a Zr content of up to ~25 at.% (after 25 hours) and a Gd content of up to ~1.5 at.% (Supplementary video 1). Fig. 2d shows the atomic-scale HAADF STEM image across the interface along the [101] zone axis, revealing that the Zr-rich regions have a superlattice-ordered $L1_2$ crystal structure with the stoichiometry of $Al_3(Zr, Gd)$, which is based on the face-centred-cubic (fcc) template structure of α-Al, forming a near-conformal dense nano-layer on the brittle $Al_3Gd$ fibres, as shown in Fig. 2e (a low magnification top view is also shown in Supplementary Fig. S4). These are termed as superlattice nano-layered (SNL) fibres.

## Formation of core-shell nano-particles in primary α-Al of Al-Gd-Zr (AGZ-SNL) alloy

Fig. 3a shows a low-magnification HAADF STEM image from the primary α-Al region of the Al-Gd-Zr alloy after 25 hours of annealing at 400°C. A high fraction of near-spherical particles with an average radius of ~2.5 nm (between 1 and 3 nm) was distributed homogeneously with an average interparticle spacing of ~25 nm in the matrix. The volume fraction is estimated to be 0.1% in α-Al and remains constant across the microstructure (Supplementary Fig. S5). The atomic-resolution image, featuring a few particles along the [001] cubic zone axis, is shown in Fig. 3b. This reveals that the particles are fully coherent with the surrounding fcc α-Al matrix and exhibit a faceted structure along the 010 and 110 planes (indicated by yellow dashed lines). The Fast Fourier Transformation (FFT) of the particles show that their crystallisation occurs in the $L1_2$ crystal structure. The EDS (Energy Dispersive Spectroscopy) atom map across one of the smaller nano-particles (~4 nm) reveals that Gd and Zr occupy the same lattice positions of {0,0,0} corners while Al occupies the {1/2,1/2,0} face-centred positions of the $L1_2$ unit cell, indicating these particles are of stoichiometry $Al_3(Zr, Gd)$. In contrast, the HAADF images of relatively larger particles (~7 nm) show high Z-contrast (atomic number contrast) concentrated at the core of the particles surrounded by a shell with an intermediate contrast, as shown in Fig. 3c. The EDS map (Supplementary S6) across these particles reveals enrichment of Gd at the core, while Zr is distributed uniformly. Fig. 3d shows an atom probe reconstruction with the distribution of a few particles in the α-Al matrix. A 2D compositional map across a larger particle shows that Gd increases to up to 4 at.% at the core, gradually decreasing towards the particle/matrix interface. These particles are termed core-shell nano-particles. Supplementary Fig. S5 also shows atom probe reconstructions from primary α-Al located near the mould wall and at the centre of the microstructure, showing a similar distribution of nano-particles, with a few exhibiting a core-shell structure, with Gd composition at the core increasing up to ~12%. Fig. 3e-f shows lattice strain maps across a core-shell particle and a homogeneous particle, revealing a high strain of ~0.04 at the core that decreases gradually towards the interface. In contrast, the high strain changes abruptly across the interface in the latter case.

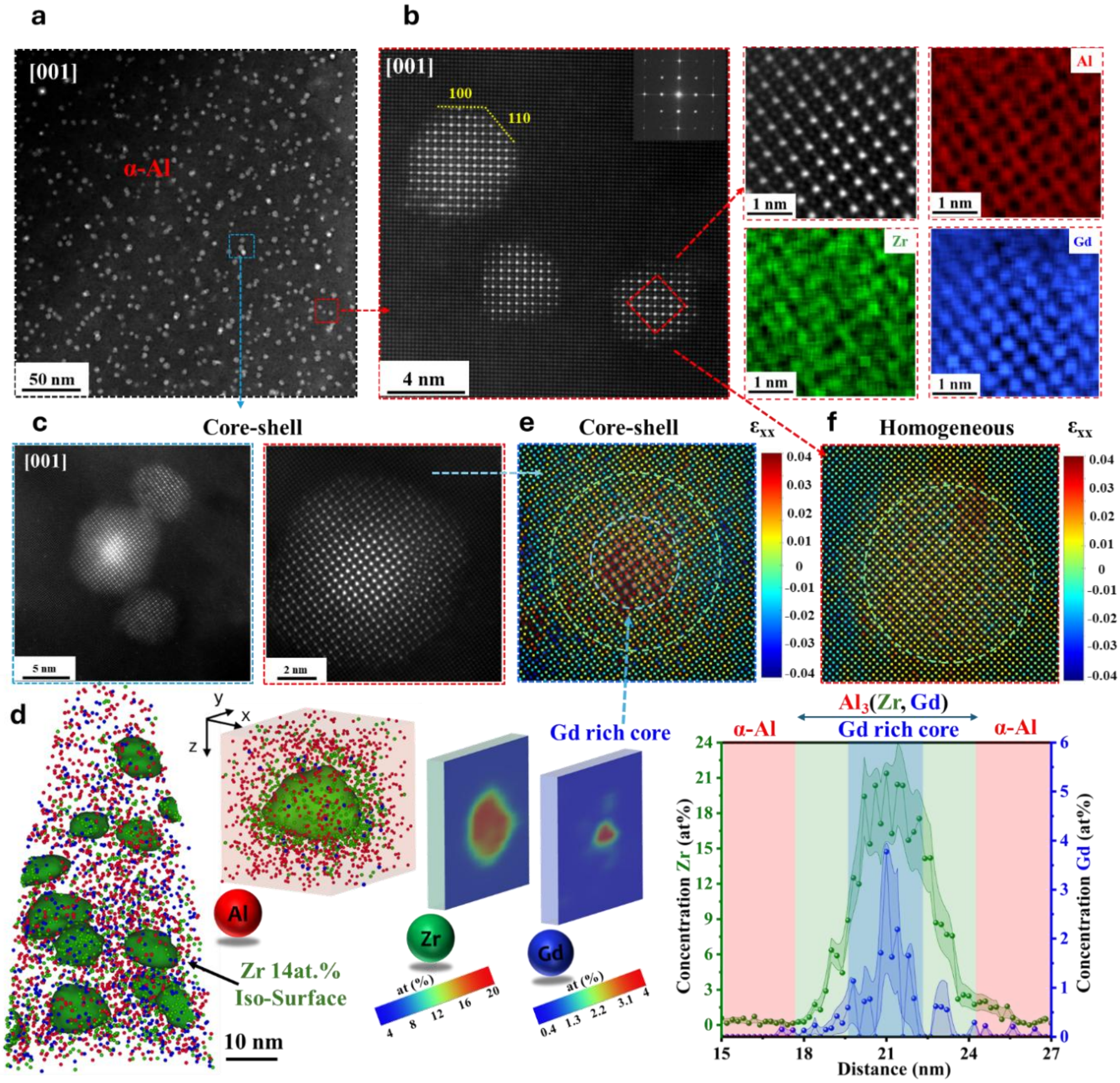

**Fig. 3: Formation of coherent strengthening nano-particles in primary α-Al of AGZ-SNL alloy: a** A low magnification HAADF STEM image from the primary α-Al region showing homogenous distribution of near spherical nano-particles. **b** Atomic-resolution HAADF STEM images showing coherent nanometric ordered nano-particles in the primary α-Al matrix. Atomic-EDS map from a nano-particle orientated in [001] cubic zone axis showing site occupancy of Al at {1/2,1/2,0} lattice positions and Zr/Gd at {0,0,0} positions of the $L1_2$ unit cell. **c** Atomic-resolution HAADF STEM images showing coherent nanometric core-shell nano-particle in the primary α-Al matrix that reveals higher Z contrast of Gd atoms at the core. **d** An APT reconstruction with the Al, Zr, and Gd atom distribution, with the iso-compositional surfaces delineated by Zr 14 at.% representing the ordered nano-particles. A 2D compositional map of Zr and Gd projected onto the yz plane shows a clear Gd composition at the core of a nano-particle. The compositional profiles across the nano-particle show an increase in Gd up to approximately 4 at.% at the core. **e-f** The strain map across a core-shell nano-particle shows a larger strain at the core. It gradually reduces towards the interface and across a nano-particle without a core, where the strain is uniform until the interface.

## Mechanical properties of Al-Gd-Zr (AGZ-SNL) alloy

Fig.4a shows representative tensile engineering stress-strain plots (all the plots are shown in Supplementary Fig. S7) for the as-cast AG and AGZ, along with the annealed AGZ having SNL fibres and core-shell particles (AGZ-SNL). The cast AG alloy fractures as it reaches necking with a tensile strength of ~205 ± 10 MPa. The tensile curve of AGZ-SNL alloy shows a fivefold increase (~400%) in ductility up to ~ 20 ± 2% as compared to AG (~ 4 ± 1%), along with a strength improvement by 50% up to 295 ± 15 MPa. In AGZ-SNL, the ductility continues even after necking, i.e., after plastic instability, unlike in AG. Fig. 4a also indicates that the cast AGZ alloy (without annealing) exhibits similar behaviour to AG. Fig. 4b presents an Ashby plot (See Supplementary Fig. S8 for the nomenclature) comparing tensile strength versus elongation for cast eutectic alloys (Al-Si, Al-Ni, Al-Ce-based, Al-Cu-Mn-based, Al-Cu-Ce-based, Al-Cu-Mg-based, and Al-Si-Mg-Cu-Ni-based) with the AGZ-SNL alloy, which exhibits an excellent combination of tensile strength and elongation. Additionally, the degree of strain hardening for AGZ-SNL alloy is higher than for AG alloy, as shown by the strain hardening rate vs. true strain plot in Supplementary Fig. S9. At 250°C, the alloy retains a strength of up to 130 ± 10 MPa, compared to a lower value for AG (90 ± 15 MPa).

Supplementary Fig. S10 shows a low-magnification HAADF STEM image of the AG alloy after fracture that reveals a large number of microscopic fracture events (indicated by red arrows) that finally lead to integral catastrophic failure across the brittle fibres. Severe debonding at the fibre/matrix interfaces is also observed, as shown in Fig. 5a. In the case of the AGZ-SNL alloy strained up to ~3% after plastic instability, Fig. 5b shows the resistance offered by the SNL against severe fracture/debonding at the α-Al/fibre interface. In the 250℃ tensile fracture sample, the SNL was also retained (Supplementary Fig. S11), indicating its stability and contribution to high-temperature strength. In the primary α-Al matrix, Fig. 5c reveals the formation of ultrafine, dense dislocation networks with sizes of ~ 12 nm.

The creep performance of AGZ-SNL is also highlighted in Supplementary Fig. S12, which shows the steady-state creep rate as a function of applied stress compared to other previously reported Al-based alloys with notable creep resistance. The AGZ-SNL alloy exhibits a creep rate nearly two orders of magnitude lower than that of the AG alloy at an applied stress of ~ 60 MPa. Compared to other alloys, the AGZ-SNL sustains a creep rate several orders of magnitude lower than most of the alloys and similar to reported Al-Ce-Sc[3] and Al-Ni-Sc[16] eutectic alloys. However, their creep rate increases rapidly with stress (stress exponent value, $n > 27$) as compared to AGZ-SNL ($n \sim 12$, see Supplementary Fig. S12a). Additionally, the Al-Ce-Sc alloy[3] has relatively poor tensile ductility, up to ~ 6%, which is 230% lower compared to the AGZ-SNL alloy.

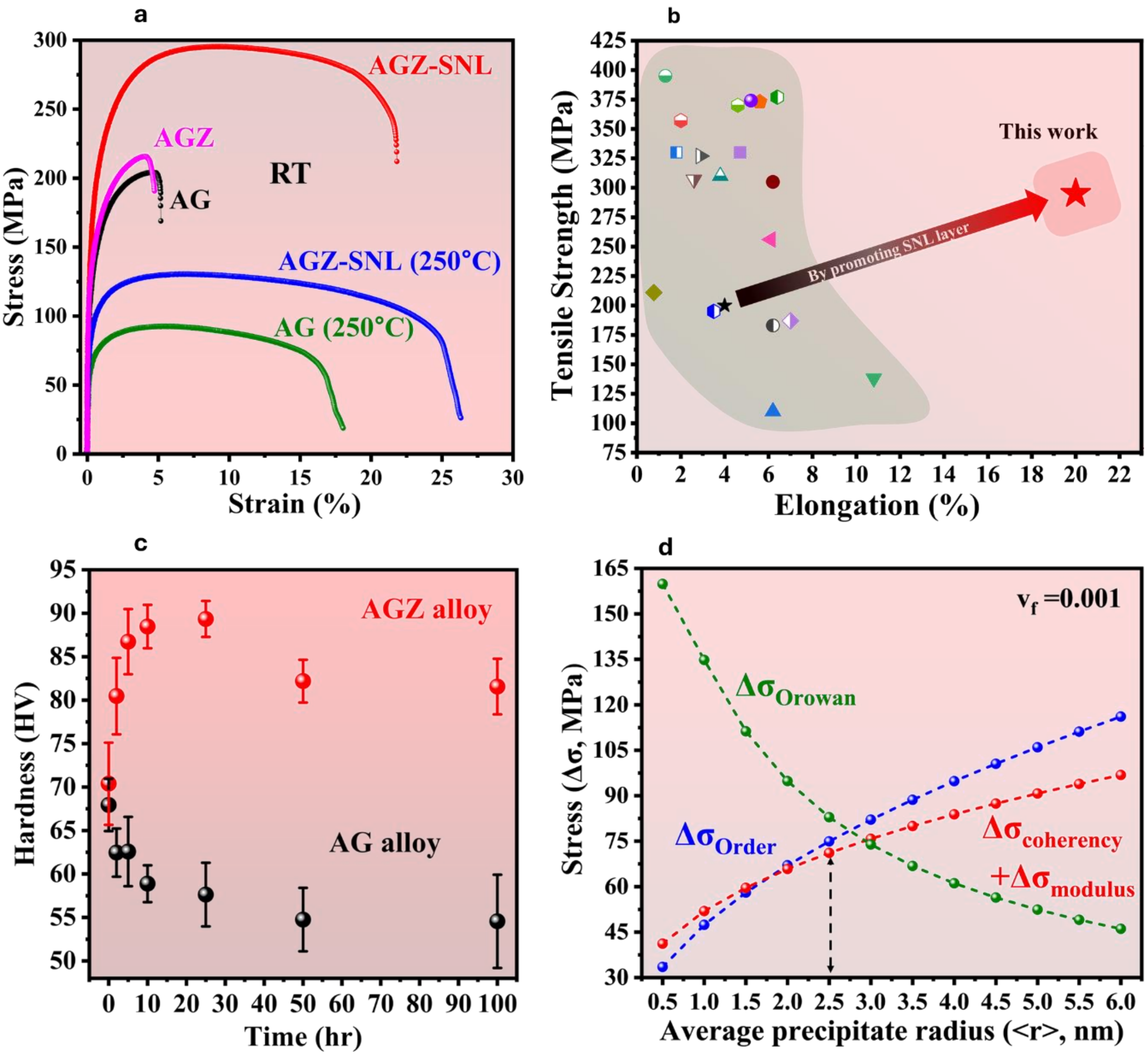

**Fig 4: Mechanical properties of Al-Gd-Zr (AGZ-SNL) alloy: a** Comparison of engineering tensile stress-strain plots at room temperature showing a fivefold increase in tensile plasticity for AGZ-SNL compared to AG alloy. At 250°C, the YS retains up to 130 ~MPa for AGZ-SNL alloy. **b** An Ashby plot between tensile strength and elongation for available high-strength cast Al eutectic alloys (Al-11.5Si[1], Al-5.8Si-0.36Sc[23], Al-2.9Ni[2], Al-2.9Ni-0.25Sc[2], Al-2.7Ce[3], Al-2.7Ce-0.3Sc[3], Al-Cu-Mn-based[24], Al-Cu-Ce-based[25,26],Al-Cu-Mg-based[27], and Al-Si-Mg-Cu-Ni-based[28–30]) in comparison to Al-2.5Gd (AG) and Al-2.5Gd-0.15Zr (AGZ-SNL) alloys. **c** The plot between the hardness and annealing time at 400°C for AG and AGZ alloy shows that the hardness drops drastically for AG, while it increases to 90 HV after 25 hours and remains at 84 HV even after 100 hours. **d** Mapping of dominant strengthening contribution as a function of the size (radius) of the nano-particles present in the primary α-Al matrix.

Microstructural stability at 400°C was evaluated for up to 100 hours through hardness measurements. Fig. 4c shows the variation in room-temperature hardness with annealing time for the AGZ and the AG alloy. The hardness rapidly increases up to 12 hours, reaching ~86 HV, followed by a plateau up to 25 hours, and then decreases to ~83 HV, which is retained even after 100 hours. The initial rapid increase in the hardness for AGZ indicates the development of SNL fibres in the eutectic structure and formation of coherent nano-particles in the primary α-Al. In contrast, in AG, a drastic drop in hardness right after the initial annealing time to a value of ~ 52 HV after 100 hours implies rapid coarsening of $Al_3Gd$ fibres. The Supplementary Fig. S13 shows the comparison of microstructural evolution in the binary (AG) and the ternary (AGZ) alloys. After 2 hours, both alloys had a few regions of coarsened eutectic structures. After 10 hours, the fraction of eutectic regions increased in both alloys, but the difference between the alloys is more evident after 25 and 100 hours; specifically, the fraction of coarsened eutectic regions (brighter contrast regions) is higher in the AZ alloy (~15% after 25 hours and ~25% after 100 hours) indicating relatively higher microstructural stability for the ternary AGZ alloy (~5% after 25 hours and ~10% after 100 hours).

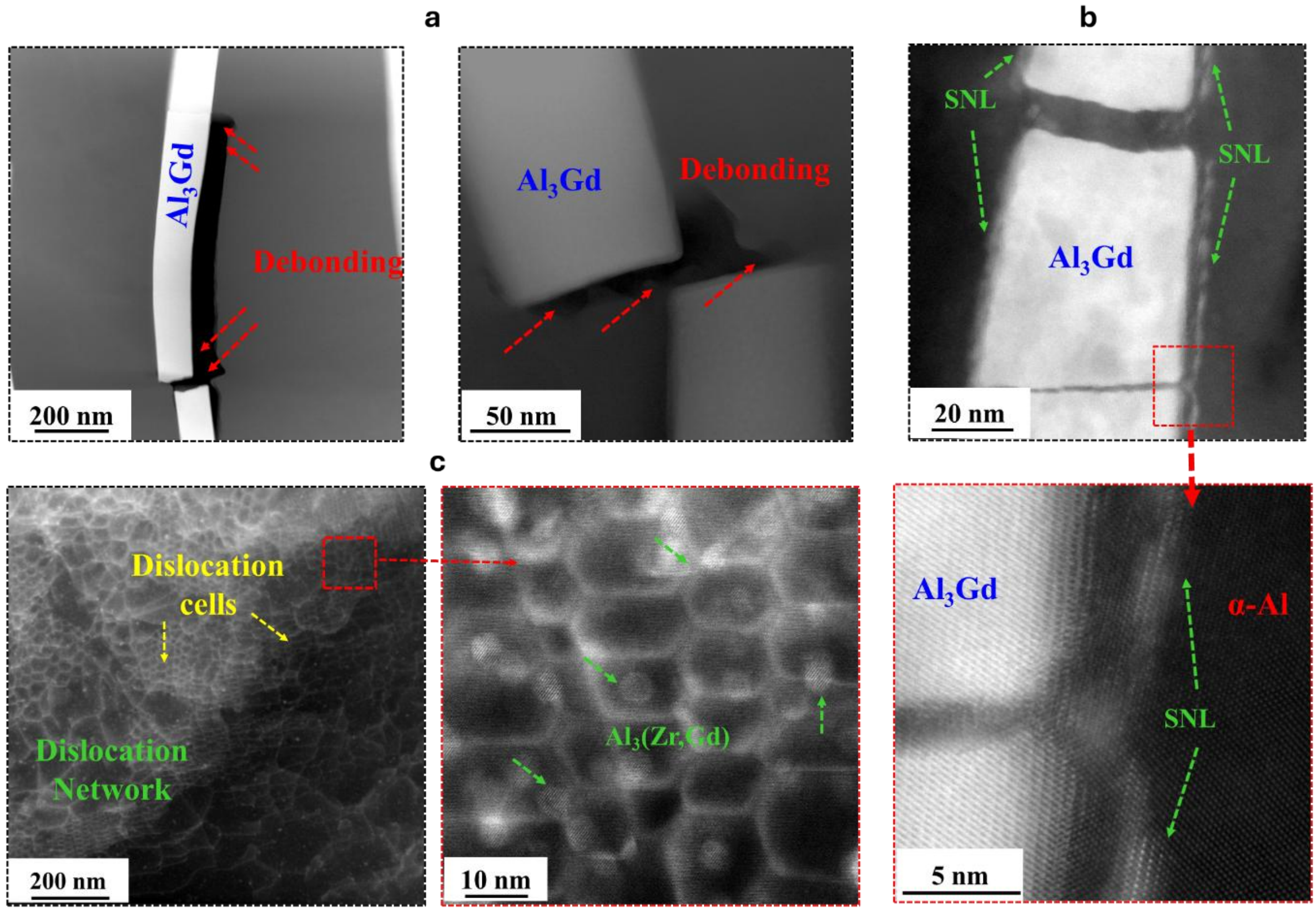

**Fig 5: Deformed microstructure of Al-Gd (AG) alloy (after fracture) and Al-Gd-Zr (AGZ-SNL) alloy with SNL fibres and nano-particles in primary α-Al (after 3% plastic strain):** **a** A HAADF STEM image centred on brittle fibre/α-Al in Al-Gd binary alloy showing severe debonding indicated by red arrows. **b** Similar HAADF STEM images of fibre/α-Al interface in Al-Gd-Zr alloy showing the resistance offered by the SNL against the severe debonding and fracture. **c** A low-AADF (LAADF) STEM image from the primary α-Al region that reveals the formation of highly dense nanoscale dislocation networks (~ 12 nm) due to the presence of coherent $Al_3(Zr, Gd)$ nano-particles.

**Discussion**

## On the development and stability of superlattice nano-layered (SNL) fibres in Al-Gd-Zr alloy

The α-Al/$Al_3Gd$ interfaces are found to be incoherent, and in a few instances, they are semi-coherent with the presence of interfacial dislocations (Fig. 1c). Fig. 6a shows an α-Al/$Al_3Gd$ interface with the presence of $Al_3$(Zr, Gd) SNL in between. The interface contains specific facets corresponding to growth ledges (as indicated by orange arrows) and, hence, along with the interfacial dislocations, they act as nucleation sites that are expected to increase the solute segregation tendency and accelerate the formation of SNL around the brittle $Al_3Gd$ fibres [31,32]. Here, the growth ledges imply that fibres thicken by a mechanism similar to plate-shaped nano-particles in numerous Al-alloys[33–35]. During annealing, the growth kinetics are restricted by the migration rate of these ledges, controlled by the diffusion of Gd towards the ledge from the matrix, as schematically shown in Fig. 6b. In the AGZ-SNL alloy, Zr atoms segregate at the fibre/matrix interfaces, thereby favouring the formation of the ordered $Al_3$(Zr, Gd) phase to minimise the overall Gibbs free energy of the system.

The merging of these individual nano-particles leads to the formation of the SNL with a lower free energy configuration, which limits the thickening rate of the fibres, as further growth requires the diffusion of Gd atoms across the SNL by breaking low-energy and stable Al-Zr/Gd atomic arrangements. This will also directly affect the microstructural stability of the alloy during annealing at 400°C. More specifically, the hardness of AG alloy continuously drops to a value of ~52 HV after 50 hours of annealing, which indicates rapid coarsening of $Al_3Gd$ fibres through Ostwald ripening[36]. This implies that the larger phase coarsens at the expense of the smaller fibres that dissolve in the matrix, partly due to the Gibbs-Thompson curvature effect[36]. Hence, in the AG alloy, coarsening occurs by the dissolution of $Al_3Gd$ fibres in the α-Al solid solution matrix, followed by diffusion of Gd atoms towards the larger $Al_3Gd$ fibres driven by the differences in the chemical potential for the Gd atoms. However, in the AGZ-SNL alloy, the diffusion of Gd atoms can be arrested by the SNL around the fibres promoted by its partitioning to the $L1_2$-ordered structure of SNL. This leads to disordered zones (DZs) between the fibre and surrounding SNL, as shown in Fig. 6c-e. The ratio of Zr:Gd in the SNL is measured to be 3:1 (See Supplementary Fig. S14), indicating a higher amount of Gd partitioning to the SNL after the formation of DZs as compared to its initial composition. Hence, the coarsening of $Al_3Gd$ fibres in AGZ alloy is expected to be lower than that of the AG alloy. From the atomic-scale HAADF image shown in Fig. 6e, the orientation relationship was found to be (Fig. 6d).

$$[10\bar{1}0]_{Al_3Gd} \,||\, [110]_{\alpha Al} \,||\, [110]_{SNL}$$

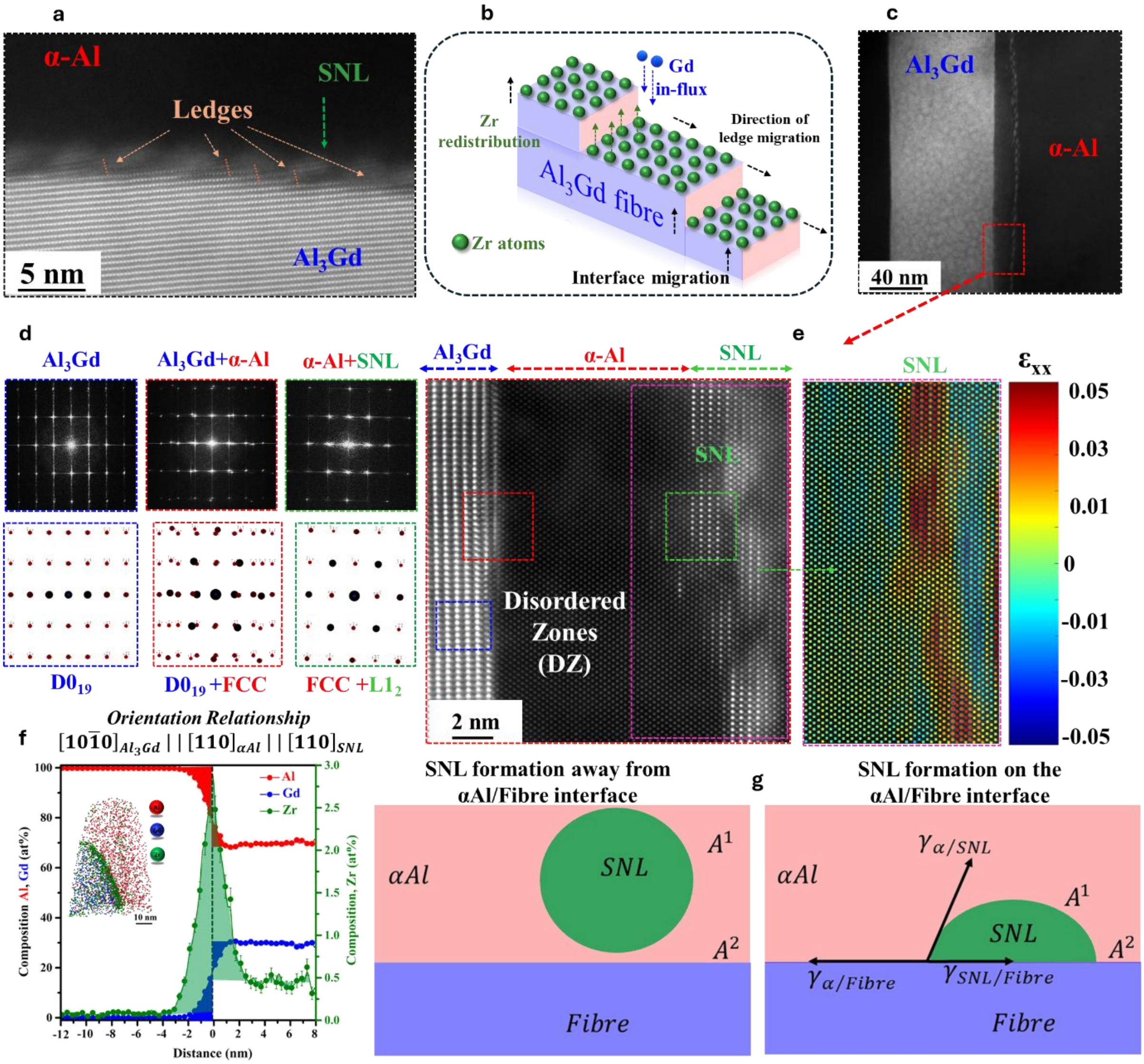

**Fig 6: Growth mechanism of SNL at the interface of brittle $Al_3Gd$ fibres in Al-Gd-Zr (AGZ-SNL) alloy: a** A HAADF STEM image taken from an $Al_3Gd$ fibre/α-Al interface in AGZ showing the presence of ledges, indicating that the fibre growth follows classical ledge mechanisms that require diffusion of Gd atoms towards the riser of ledges from the matrix and redistribution of Zr atoms, as shown also schematically in **b. c** HAADF STEM images centred on an $Al_3Gd$ interface show a disordered zone (DZ) between the fibre and SNL. **d** Orientation relationship between the $Al_3Gd$/α-Al/SNL. **e** The strain map across the SNL shows a larger strain along the layer with respect to the matrix. **f** The Gibbsian interfacial excess quantities of Al, Zr and Gd shown as shaded areas under the compositional curves in the proximity histogram across an αAl/$Al_3Gd$ fibre interface after 5 hours of annealing. **g** Schematic representation of the tendency of formation of the SNL layer on the $Al_3Gd$ fibres in α-Al.

Here, two distinct driving forces may govern the formation of the ordered $Al_3Zr$ phase at the α-Al/$Al_3Gd$ interfaces. First, the interfaces can act as preferential heterogeneous nucleation sites for the $Al_3Zr$ phase. Second, there exists a thermodynamic driving force for Zr segregation to the α-Al/$Al_3Gd$ interfaces. Based on our APT results, which indicate that Zr segregation occurs prior to the formation of the $Al_3Zr$ phase at short ageing times (see Fig. 2a–c), we infer that the latter mechanism is the dominant driving force. The interfacial segregation around the brittle $Al_3Gd$ fibres can be understood by the quantitative analysis of the concentration of solute atoms (Zr) across a heterophase interface (αAl/$Al_3Gd$), and rationalised based on the relative Gibbsian interfacial excess using a thermodynamic framework[37,38] described in the Methods section (Appendix A). Fig. 6f highlights the Gibbsian interfacial excess for Al, Zr and Gd as shaded areas under the compositional profiles in the proximity histogram. The relative Gibbsian interfacial excess of Zr with respect to Al and Gd is found to be 3.25 atoms/$nm^2$. The decrease in the interfacial free energy associated with Zr segregation is estimated to be ~ -110 $mJm^{-2}$. This decrease is critical, since most of the α-Al/$Al_3Gd$ interfaces are incoherent, which have high interfacial energy but no associated elastic strain energy.

The thermodynamic driving force for SNL formation can be rationalised using classical nucleation theory[36]. Fig. 6g shows a schematic representation of the tendency of formation of the SNL layer on the $Al_3Gd$ fibres in α-Al. For homogeneous nucleation of SNL in the α-Al matrix in the presence of $Al_3Gd$ fibre, the total system Gibbs free energy ($\Delta G_{hom}$) is given by

$$\Delta G_{hom} = -V\Delta G_v + \Delta G_s^1 + (\gamma_{\alpha Al/fibre} \times A^2_{\alpha Al/fibre}) + (\gamma_{\alpha Al/SNL} \times A^1_{\alpha Al/SNL}) \qquad (1)$$

Where, $V\Delta G_v$ is the total volume-free energy reduction due to the precipitation of SNL in α-Al, which contributes to reducing the total system's free energy. $\Delta G_s^1$ is the strain energy including contributions from both SNL/α-Al and fibre/α-Al interfaces. $\gamma_{\alpha/Fibre}$ and $\gamma_{\alpha/SNL}$ are the interfacial energies per unit area of α-Al/fibre and α-Al/SNL interfaces, while $A^1$ and $A^2$ are the respective surface areas of the interfaces as shown schematically in Fig. 6g. Note that the last three terms tend to increase the total system's free energy.

The formation of SNL at the interface of α-Al/$Al_3Gd$ fibres is driven by the reduction in the total system's free energy ($\Delta G_{het}$) due to the contribution from the minimisation of 1) interfacial energy, and 2) misfit strain energy, which can be understood by the following equation:

$$\Delta G_{het} = -V\Delta G_v + \Delta G_s^2 + (\gamma_{\alpha Al/fibre} \times (A^2_{\alpha Al/fibre} - A_{flat})) + (\gamma_{\alpha Al/SNL} \times (A^1_{\alpha Al/SNL} - A_{curve})) + (\gamma_{SNL/fibre} \times A_{flat}) \qquad (2)$$

The reduction in total interfacial energy occurs by the contribution due to the annihilation of surface areas of $A^2$ and $A^1$ due to the formation of SNL at the α-Al/fibre interface. The last term in equation (2) corresponds to the interfacial free energy per unit area of the new interface $\gamma_{SNL/fibre}$. The positive contribution from this surface is negligible compared to $\gamma_{\alpha Al/fibre}$ which represents for an incoherent high interfacial energy interface. By comparing equations (1) and (2), $\Delta G^{int}_{hom} \gg \Delta G^{int}_{het}$. Hence, the system will prefer the formation of SNL on the interfaces of αAl/$Al_3Gd$ fibre in the microstructure.

The misfit cross the α-Al/SNL interface is calculated as $\delta_{\alpha Al/SNL} \sim +0.87\%$ (see Appendix A in Methods). As previously mentioned, most of the α-Al/$Al_3Gd$ interfaces are incoherent. Some segments are semi-coherent with misfit strain relieved by arrangement of dislocations, as shown in Fig. 1c. The unconstrained misfit across the α-Al/$Al_3Gd$ interface is $\delta_{\alpha Al/Fibre} \sim$ +7.06%. To fully relieve the associated strain, the minimum required distance between the interfacial dislocations is estimated to be ~ 2.16 nm. From Fig. 1c, the distance between the two dislocations ~5.1 nm, indicating the interface is still under misfit strain. Thus, the experimentally determined misfit strain, $\delta^{exp}_{\alpha Al/Al_3Gd}$ is estimated to be ~ +2.99% for the semicoherent interface segments of the interface. The misfit strain energy $(\Delta G_{strain})$ for both α-Al/$Al_3$Gd and α-Al/SNL interfaces are found to be ~ +93 $MJ/m^3$ and ~ +7.8 $MJ/m^3$, respectively. Hence, the localised precipitation is also driven by minimising the misfit strain energy[39] across α-Al/$Al_3$Gd fibre by promoting SNL at the highly strained interfaces, reducing the effective misfit strain to ~ +6.14% or $\Delta G_{strain}$ ~ +85.2 $MJ/m^3$. Recently, A similar concept of segregation-assisted precipitation was exploited to form a phase around larger nano-particles having misfit dislocations that exhibits high H-storage capacity for making H-resistant Al alloys[40].

## On the development of core-shell nano-particles in Al-Gd-Zr alloy

In α-Al, the finer nano-particles (2 to 4 nm) have a uniform Gd composition up to ~2 at.%, whereas the larger size nano-particles exhibit Gd enrichment at the core (up to ~4 at.%) that gradually reduces towards the particle/matrix interface. The atomic size of Gd is 160 pm (larger by ~ 14% with respect to Zr), that comes in the rare-earth (RE) lanthanide series of elements in the periodic table, which are known to expand the unit cell of $L1_2$ $Al_3$X (X-Zr, Sc) by occupying corner {0,0,0} positions[41–43]. This led to an increase in the lattice misfit strain at the interface. Upon annealing, during the growth of the nano-particles, the Gd diffuses to the core, driven by the reduction in the elastic strain energy of the interface, as confirmed by lattice strain maps[44] across the nano-particles without and with the core in Fig. 3e-f. In the former case, a high strain of ~0.04 in the nano-particle changes abruptly across the interface toward the matrix. In the core-shell nano-particle, we see a higher strain at the core, which reduces gradually towards the interface, showcasing how the diffusion of Gd modulates the misfit from the core to the interface.

## On the increased tensile plasticity of Al-Gd-Zr alloy (AGZ-SNL)

In the microstructure, the initial plastic deformation occurs in the FCC primary α-Al and α-Al inside the eutectic structure, generating many dislocations that glide on the primary {111} planes. Due to the high stacking fault energy of Al (~ 166 $mJ/m^2$)[45], it is expected that dislocations will easily cross-slip across several {111} planes upon loading, which should directly lead to high ductility. However, in AG and AGZ alloys with an area fraction of α-Al of approximately 45%, the microstructure fails catastrophically before necking. In the AGZ-SNL alloy, plastic strain accumulation continues even after necking, reaching ~20% before fracture, indicating the importance of the modified eutectic structure by the SNL fibres. In the binary alloy (AG), as shown in Supplementary Fig. S10, the fractured sample reveals multiple cracks propagating simultaneously across the brittle $Al_3$Gd fibres, indicating that sudden fracture occurs immediately after necking and leads to plastic instability during the uniaxial tensile test. This can be attributed to the development of high-stress concentration sites at the interfaces upon loading, facilitating crack nucleation and severe debonding (Fig. 5a), which releases stress that the surrounding α-Al[10,11] cannot absorb.

To reveal the deformation behaviour in the AGZ-SNL alloy, a sample with 3% plastic strain (after plastic instability) was imaged using low-angle annular darkfield (LAADF) mode, which provides strain contrast. Fig. 7a shows a low-magnification LAADF image centred on an α-Al region between the two SNL fibres. Fig. 7b is a close-up on the region indicated by a black dashed rectangle in Fig. 7a. Dislocations are seen to accumulate behind the SNL and appear pinned at the matrix/SNL interface, indicating resistance to shear in the layer, as shown in Fig. 5b-c. This can be attributed to the high local strain of ≈ 0.05% relative to the surrounding α-Al phase (Fig. 6e), which prevents the easy glide of the dislocations. This leads to the accumulation and intersection of dislocations, which form networks behind the impenetrable layer, as observed in Fig. 7b-c. The compositional mapping (supplementary Fig. S15) shows that the Zr/Gd is confined only to the SNL, not at the dislocations behind. The atomic arrangement of the dislocations is shown in Fig. 7d, indicated by yellow dashed circles, revealing that the dislocations do not glide on a single plane, suggesting a high tendency to form a dislocation network.

In the case of binary AG alloys, the α-Al/$Al_3Gd$ interfaces are not coherent, i.e., the crystal lattices are not continuous across them. Hence, dislocation transmission into the fibres is hindered by the need to change slip direction and by changes in the magnitude of the Burgers vectors. Additionally, the fibres are an HCP-based ordered structure ($DO_{19}$) with limited slip systems[46], requiring relatively higher stresses for activation than for the softer FCC-based $L1_2$ ordered structure. As a consequence, the accumulation of dislocations at $Al_3Gd$/Al interfaces and associated stress concentration leads to extensive debonding at these interfaces. In ternary AGZ alloys, the ordered SNL phase covers the interface, and dislocations cannot reach the brittle $Al_3Gd$ fibre by cross-slipping or looping. The SNL is fully coherent with the α-Al, characterised by continuous crystal lattices with their orientation of the ordered $L1_2$ phase identical to that of the disordered FCC α-Al phase[47]. We observe that dislocations do not transmit into the ordered SNL phase (Fig. 7b-c), possibly because of the increment in transmission stress required due to APB creation (Fig. 4d). Thus, in the AGZ alloy, dislocation accumulation will occur at the $Al_3Zr$ SNL phase rather than at the $Al_3Gd$/Al interface. Moreover, the SNL phase may accommodate the resulting stress by plastic deformation, as it is a relatively ductile $L1_2$ phase[48]. The α-Al/SNL interface, due to its coherent nature, is likely to be much stronger than the incoherent/semicoherent α-Al/$Al_3Gd$ interface, thus reducing the tendency for debonding as is observed in the ternary AGZ alloy (Fig. 4). We believe that this unexpected interface engineering is the main cause of the increase in ductility at higher strength levels in the ternary AGZ alloy. In AG alloy, the stress concentration sites increase the propensity of crack initiation events, and their propagation cannot be effectively buffered and arrested, which leads to catastrophic failure before the onset of necking. Similar shielding against high local stresses, leading to ductility upto ~ 50%, is also observed in the directionally solidified Ni-Co-Fe-Al-based eutectic high entropy alloy that contains a lamellar microstructure of alternate hard B2 ordered and soft $L1_2$ ordered phases[49]. Here, the softer $L1_2$ serves as a crack buffer, blunting them and shielding the associated high local stresses.

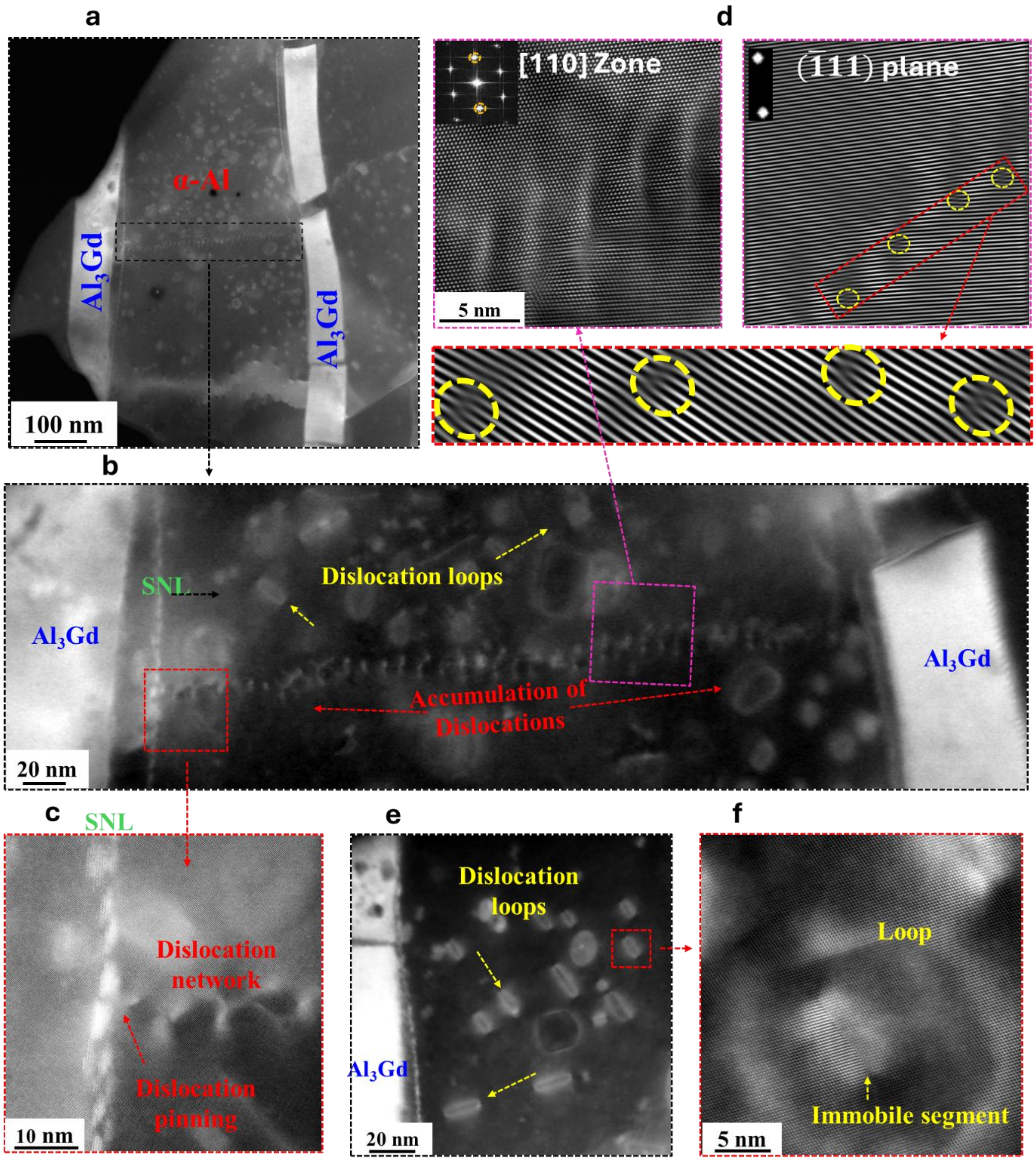

**Fig. 7: Role of SNL on the operative deformation mechanisms in α-Al during tensile loading of AGZ-SNL alloy: a** A low magnification HAADF STEM image centred on α-Al in the eutectic region of AGZ alloy deformed 3% after plastic instability, showing an accumulation of dislocations behind the SNL in α-Al **b-c** preventing them from reaching the weak eutectic α-Al/$Al_3Gd$ fibre interface. **d** A high-resolution HAADF STEM image taken centred on a few accumulated dislocations (pink dashed square) showing the dislocations on different planes (yellow circled). **e-f** LAADF STEM images show strain contrast in the α-Al of the eutectic region, which is attributed to the presence of a large number of dislocation loops and immobile segments, indicating the occurrence of multiple/cross slips during plastic deformation.

The α-Al phase also contains a high density of dislocation loops, Fig. 7e, generated continuously around the immobile dislocation, so-called sessile jogs, restricting dislocation motion and activating Frank-Read sources[50,51] that nucleate new dislocations during plastic deformation, leading to progressive work hardening of the alloy. Fig. 7f shows the atomic structure across one such source centred on an immobile dislocation segment and a dislocation loop around it. These loops again encounter obstacles during slip, resulting in restricted motion of the dislocation lines. These act jointly with dislocation networks as hard obstacles in α-Al, and plastic deformation will proceed by continuous extrusion of dislocation between the sessile jogs and the network walls, like the operation of a Frank-Read Source, contributing to the higher strain hardening rate for the AGZ-SNL alloy compared to the AG alloy (supplementary Fig. S9).

In primary α-Al, the formation of a dense dislocation network (~ 12 nm) is the result of extensive dislocation activity through a large number of dislocation cross-slip/multiple-slip events across different {111} planes[52] in the presence of coherent nano-particles (see Supplementary Fig. S16 for elemental map across the network). These events lead to a high degree of dislocation, intersections and entanglement. As the plastic strain increases, these entangled dislocations distribute heterogeneously into more stable, lower-energy networks.

The dislocation density ($\rho$) in the soft α-Al matrix for both tensile deformed AG and AGZ-SNL alloy was estimated from the LAADF STEM images (Supplementary Fig. S17) containing dislocation cells by using the formula[53]:

$$\rho = \frac{1}{(cell\ size)^2} \tag{3}$$

The local shear stress ($\tau$) in α-Al matrix is estimated by the Taylor formula:

$$\tau = \alpha G_{Al} b \sqrt{\rho} \tag{4}$$

Where α is constant (~ 0.3 for Al), $G_{Al}$ is the Shear modulus of Al (~25.4 GPA at 25°C)[54], and $b$ is the Burgers vector (0.286 nm for Al)[54].

Based on the network's cell size distribution, the total dislocation density is estimated to be ~ $6.4 \times 10^{15}$ $m^{-2}$ in the primary α-Al matrix, a one order of magnitude higher than the binary AG alloy after tensile fracture (~ $1.8 \times 10^{14}$ $m^{-2}$). This difference led to an over five-fold increase in the shear strength of α-Al (~ 160 MPa) from ~30 MPa for the AG alloy. Here, the increase in the dislocation storage capacity of the α-Al matrix can be attributed to (i) extensive pinning of dislocation glide promoting more cross-slip events, (ii) higher strength, and (iii) the delay in fracture for the AGZ-SNL. The inability of dislocations to shear the nano-particles is associated with the difference in the elastic modulus and high misfit strain across the coherent interfaces between nano-particles[55,56] and the matrix.

The strengthening contribution from the nano-particles can be understood by the mechanisms that depend on their radius. These are 1) coherency strengthening, 2) modulus strengthening, 3) order strengthening, and 4) Orowan looping. Coherency strengthening is due to the interaction of dislocations with the elastically strained matrix that is related to the high misfit at the particle/matrix interface. The respective contributions are quantified based on the equation shown in Appendix 2 (see methods). Fig 4d shows the plots of the individual contributions as a function of the particle's radius. In the AGZ-SNL alloy, the average nano-particle radius is ~ 2.5 nm (ranges from 1 to 3 nm), indicating that strengthening is dominated by either order ( nano-particles with radii from 1 to 2 nm) or coherency and modulus mismatch.

To estimate the strengthening contributions quantitatively, the binary AG alloy was also tensile-tested after annealing for 25 hours to obtain mechanical properties under the same heat-treated condition as for the AGZ alloy. Supplementary Fig S7b provides the tensile curves. The measured 0.2% yield strength, $\sigma_{AG25}$, of AG alloy after 25 hours of annealing shows a value of ~49 MPa (See Supplementary Fig. S7b). The strength of the ternary alloy after a similar heat treatment is 190 MPa.

We observe that the volume fraction of the eutectic structure remains essentially unchanged (~0.55) upon Zr addition to the AG alloy. Assuming no scaling effects associated with Zr addition and considering ideal behaviour for the purpose of estimation:

The yield strength of the binary AG alloy is given by

$$\sigma_{total}^{B} = \vartheta_f^E \sigma_E + (1 - \vartheta_f^E)\sigma_{\alpha Al} \tag{5}$$

where, $\sigma_E$ and $\sigma_{\alpha Al}$ are the strengthening contributions due to the eutectic structure and α-Al matrix, and $\vartheta_f^E$ is the volume fraction of the eutectic structure.

Similarly, the yield strength of the ternary AGZ-SNL alloy is

$$\sigma_{total}^{T} = \vartheta_f^E \sigma_E^{SNL} + (1 - \vartheta_f^E)\sigma_{\alpha Al}^{NP} \tag{6}$$

where, $\sigma_E^{SNL}$ and $\sigma_{\alpha Al}^{NP}$ are the strengthening contributions due to the eutectic structure having SNL fibres and α-Al matrix with nano-particles.

The increment in the strength after Zr addition to AG alloy, followed by annealing at 400 °C for 25 hours:

$$\Delta\sigma_{total}^{T} = \vartheta_f^E \Delta\sigma_E^{SNL} + (1 - \vartheta_f^E)\Delta\sigma_{\alpha Al}^{NP} \tag{7}$$

We estimated $\Delta\sigma_{\alpha Al}^{NP}$ to be ~ 72 MPa from Fig. 4d. The value for $\Delta\sigma_{total}^{T}$ is ~141 MPa (190 MPa – 49 MPa). Since $\vartheta_f^E$ is measured to be ~0.55, the strengthening contribution from eutectic structure with SNL fibres in AGZ-SNL alloy ($\vartheta_f^E \Delta\sigma_E^{SNL}$) is estimated to be ~ 108 MPa.

In conclusion, we have demonstrated an approach based on the targeted reduction of the interfacial energy and misfit strain energy in a eutectic Al-alloy across the brittle fibre/soft matrix interface by promoting the formation of SNL fibres. We strategically exploited the addition of Zr to the Al-Gd alloy, facilitating interfacial segregation and the subsequent precipitation of an ordered nanolayer phase to counteract the detrimental stress concentration at the high-energy interfaces. This prevents the catastrophic failure upon loading from the stress concentration at these interphase interfaces, and with the additional distribution of homogenous and core-shell nano-particles in the primary α-Al matrix, allows for an exceptional increase in the tensile ductility by ~ 400%, retaining a strength of ~ 130 MPa at 250°C, and a microstructural stability over long time annealing that provides outstanding creep properties. The energetic stabilisation of the microstructure enables long-term use at high temperatures. Hence, adjusting the alloy composition allows for controlling the fraction of SNL fibres and α-Al, providing scope for a new generation of lightweight, high-strength eutectic alloys for applications in extreme conditions.

## Methods

### Alloy preparation and heat treatment

Al-2.5Gd and Al-2.5Gd-0.15Zr (all in atomic %) alloys were prepared using a vacuum arc melting unit under an argon atmosphere. The bulk ingots were cast using a water-cooled Cu mould as 6mm x 10mm x 100mm slabs. "The cast alloys were heat-treated at 400°C, followed by quenching. Supplementary Fig. S19 shows the collage of images taken near the mould wall to the centre of the cast heat-treated slab after quenching. We observe that the microstructure remains similar as we move towards centre of the slab. Further the grain size remains similar (near the mould wall: 37±13 µm, centre: 40±19 µm).

### Measurement of mechanical properties (Tensile and Creep properties)

Flat dog-bone-shaped tensile samples with a gauge length of 6mm, width of 2mm, and thickness of 1mm were cut from cast and heat-treated alloys with 6 mm gauge length, 2 mm width, and 1 mm thickness using an electric discharge machine (EDM). The hardness of the as-cast and aged samples was measured using a Vickers microhardness tester with a load of 200 grams. The tensile tests at room temperature and at 250°C were performed in (Zwick Z100) with a load cell of 10kN and operated at a constant crosshead displacement rate of $10^{-3}$ $s^{-1}$. The DIC (Digital Image Correlation) method was used to obtain accurate, precise strain measurements during tensile tests. The gauge section of the samples was coated with a speckle pattern, and images were processed using VIC-2D software to perform full-field strain analysis. A subset size of 160 × 160 $\mu m^2$ with a step size of 7 pixels was selected, providing a spatial resolution of approximately 70–75 µm. The true strain rate calculated from DIC analysis was approximately $0.95 \times 10^{-3}$ $s^{-1}$, closely matching the crosshead displacement rate set for these tests.

Creep experiments were conducted in compression at 300°C (applied stresses from 45 MPa to 80 MPa) using a lever-arm creep testing machine with a lever-to-arm ratio of 20:1. Cylindrical specimens were cut from the cast slabs using EDM with a diameter of 3 mm and an aspect ratio greater than 1.6, in accordance with the ASTM standards. The specimens were positioned between two silicon nitride plates coated with boron nitride to minimise friction and ensure an unaltered stress state during testing. A furnace with a temperature stability of ±1 °C was employed to maintain consistent thermal conditions. Strain measurements were recorded using a contact-based high-temperature extensometer with a resolution of 1 µm. Upon completion of the experiments, the specimens were cooled under load to preserve the microstructure for post-test analysis.

**Microstructural investigation by electron microscope (SEM/TEM) and atom probe tomography (APT)**

For bulk microstructural imaging, samples were cut from the as-cast and heat-treated alloys and polished using Si-grit papers (1000 to 3000 grit size), followed by cloth polishing using colloidal silica solution. Samples for TEM analysis were prepared from the heat-treated and deformed alloys using $Ar^+$ ions in a PIPS (Precision Ion Polishing System). TEM was conducted using a T20 Tecnai (Thermo Fisher Scientific) instrument operated at 200 kV. High-resolution (HR) imaging was performed using an aberration-corrected TEM (Titan Themis, ThermoFisher) in scanning transmission electron microscopy (STEM) mode, operated at 300 kV. The drift-compensated images were recorded in Velox software using a high-angle-annular-dark-field (HAADF) detector at a camera length of 160 mm and 300 mm (LAADF imaging).

The site-specific APT needles were prepared for APT experiments using a dual-beam SEM/focused-ion-beam (FIB) instrument (Scios Nanolab, Thermo Fisher Scientific)[57]. The cut lamella containing ROI was attached to the posts of the Si coupon. The lamella was sharpened by using $Ga^+$ ions accelerated at 30kV with currents from 0.5 nA to 0.1 nA, followed by a final cleaning at 2kV to remove the $Ga^+$ damage regions. The needles were field-evaporated in an atom probe (LEAP 5000XR, Cameca Instruments Inc.) operated in laser pulsing mode. A laser pulse repetition rate of 125kHz and pulse energy of 55 pJ were used. The base temperature of the needles was kept at 70K, and the target detection rate was set to 0.5%. Data analysis was conducted using the IVAS™ 3.8.10 software package.

**Appendix A**

***Estimation of Gibbsian interfacial excess***

$$\Gamma_{Zr}^{rel} = \Gamma_{Zr} - \Gamma_{Gd} \times \left(\frac{x_{Al}^{\alpha} x_{Zr}^{\alpha'} - x_{Al}^{\alpha'} x_{Zr}^{\alpha}}{x_{Al}^{\alpha} x_{Gd}^{\alpha'} - x_{Al}^{\alpha'} x_{Gd}^{\alpha}}\right) - \Gamma_{Al} \times \left(\frac{x_{Zr}^{\alpha} x_{Gd}^{\alpha'} - x_{Zr}^{\alpha'} x_{Gd}^{\alpha}}{x_{Al}^{\alpha} x_{Gd}^{\alpha'} - x_{Al}^{\alpha'} x_{Gd}^{\alpha}}\right) \quad (8)$$

where, $\Gamma_{Al}$, $\Gamma_{Gd}$, and $\Gamma_{Zr}$ are the Gibbsian interfacial excess of Al, Gd, and Zr, respectively, and $x_j^{\alpha}$ and $x_j^{\alpha'}$ are the concentrations of the components $j$ ($j$=Al, Gd, Zr) in phase $\alpha$ (α–Al) and phase $\alpha'$ ($Al_3Gd$).

The decrease in the interfacial free energy associated with Zr segregation can be estimated using the equation 9:

$$\Gamma_{Zr}^{rel} = -\left(\frac{\partial \gamma}{\partial \mu_{Zr}}\right)_{T,P} \quad (9)$$

Where, $\Gamma_{Zr}^{rel}$ is the relative Gibbsian excess of Zr with respect to Al and Gd, and $\mu$ is the chemical potential of Zr. Assuming an ideal solid solution, the chemical potential is given by $\mu_{Zr} = \mu_o + k_B T \ln(x_{Zr}^{\alpha})$ and it yields the following expression:

$$\left(\frac{\partial \gamma}{\partial x_{Zr}}\right)_{T,P} = -\frac{\Gamma_{Zr}^{rel} k_B T}{x_{Al}^{\alpha}} \quad (10)$$

Integrating the above equation 8 from 0 to the peak concentration at the interface, we obtain a decrease in interfacial free energy of -110 $mJm^{-2}$.

***Estimation of misfit strain energy*** $(\Delta G_s)$

The misfit strain energy $(\Delta G_s)$ and for a coherent phase in the matrix is given by[36]

$$\Delta G_s = 2G_{Al}\left(\frac{1+\vartheta}{1-\vartheta}\right)\delta^2 \tag{11}$$

Where $G_{Al}$ is the shear modulus of the matrix (25.4GPa for Al), ϑ is the Poisson's ratio of the matrix (0.345 for Al) and δ is the misfit strain across the matrix and phase at room temperature and was calculated using the formula:

$$\delta = \frac{d_{phase} - d_{\alpha Al}}{d_{\alpha Al}} \tag{12}$$

Here, the $d_{phase}$ and $d_{\alpha Al}$ are the unstressed interplanar spacings of matching planes of coherent phase and the matrix, respectively. From Fig. 1c, the α-Al/fibre interface is semicoherent with the presence of dislocations at every ~5.1 nm. The orientation relationship indicates $(\bar{2}4\bar{2}0)$ of $Al_3Gd$ is parallel to (220) of α-Al matrix. The respective unstressed interplanar spacings are 0.153 nm $(d_{Al_3Gd})$ and 0.143 nm $(d_{\alpha Al})$. The unconstrained misfit strain estimated by using equation 10 is $\delta_{\alpha Al/Al_3Gd}$ ~ +0.0706 (~7.06%). The minimum required dislocation spacing at the interface to relieve the misfit strain is calculated by the following equation 13[36]:

$$D_{cal} = \frac{d_{Al_3Gd}}{\delta_{\alpha Al/Al_3Gd}} = \frac{0.153}{0.0706} = 2.17\ nm \tag{13}$$

The experimentally measured dislocation spacing $(D_{exp})$ from Fig 1c is ~5.1 nm. Hence, the actual misfit strain for the semicoherent interface $(\delta^{exp}_{\alpha Al/Al_3Gd})$ is estimated to be ~+0.0299 (+2.99%) by the following equation 14:

$$D_{cal} \times \delta_{\alpha Al/Al_3Gd} = D_{exp} \times \delta^{exp}_{\alpha Al/Al_3Gd} \tag{14}$$

**Appendix B**

***Estimation of strengthening contributions***

Coherency strengthening is given by the equation[58]

$$\Delta\sigma_{coherency} = M\chi\,(\varepsilon\, G_{Al})^{3/2}\left(\frac{2\langle r\rangle v_f}{G_{Al}\, b}\right)^{1/2} \tag{15}$$

Here, M is the Taylor factor (3.06 for the Al matrix)[59], χ is a constant with the value 2.6[58], $G_{Al}$ is the shear modulus of Al (25.4 GPa at 25°C)[54], ⟨r⟩ is the average particle radius, $v_f$ is the volume fraction of the nano-particles (0.1%), b is the Burgers vector for Al (0.286 nm), and ε is the strain in the matrix given by

$$\varepsilon = |\delta|\left[1 + \frac{2G_{Al}(1-2\vartheta_{Al_3Zr})}{G_{Al_3Zr}(1+\vartheta_{Al_3Zr})}\right] \tag{16}$$

where, δ is the lattice misfit between particle and the matrix (Assuming for $L1_2$ $Al_3Zr$, ~0.75%)[60].

The modulus strengthening can be estimated by the equation[58]

$$\Delta\sigma_{modulus} = 0.0055\, M\, b\, (\Delta G)^{3/2} \left(\frac{2v_f}{G_{Al}\, b^2}\right)^{1/2} \left(\frac{\langle r\rangle}{b}\right)^{\left(\frac{3m}{2}-1\right)} \tag{17}$$

Where, ΔG is the difference in shear modulus of the nano-particle and Al (Assuming $Al_3Zr$, the value is 31.6 GPa), and m is the constant (0.85)[58].

The order strengthening is given by (independent of particle size)

$$\Delta\sigma_{order} = \frac{M\, \gamma_{APB}^{Al_3Zr}}{2\, b}\left(\frac{3\, \pi^2 \gamma_{APB}^{Al_3Zr}\, v_f <r>}{16\, G_{Al} b^2}\right)^{1/2} \tag{18}$$

$\gamma_{APB}^{Al_3Zr}$ is the anti-phase boundary energy (assuming for $Al_3Zr$, 0.445 $Jm^{-2}$)[61].

For a larger particle size, the Orowan strengthening is given by

$$\Delta\sigma_{orowan} = \left(\frac{0.4\, M\, G_{Al}\, b}{\pi\sqrt{1-\vartheta}}\right)\left(\frac{1}{\lambda}\right)\left(ln\frac{\pi\langle r\rangle}{2b}\right) \tag{19}$$

Where, $\vartheta$ is the Poisson's ratio (0.345)[59] and λ is inter-particle distance and is estimated by the equation

$$\lambda = \langle r\rangle\left(\sqrt{\frac{2\pi}{3V_f}} - \frac{\pi}{2}\right) \tag{20}$$

**Acknowledgments**

We gratefully acknowledge the **Advanced Facility for Microscopy and Microanalysis (AFMM)** at the Indian Institute of Science, Bangalore, for providing access to TEM and APT facilities. The cast alloys were ingeniously developed in the **Alloy Design Laboratory (ADL)** at the **Department of Materials Engineering, IISc Bangalore**. All the Microscopy and Tomography experiments were conducted at **AFMM, IISc Bangalore**. **S.K.M.** acknowledges the financial support from the **SERB (ANRF) SRG Grant, DIA-RCOE at IISc, Bangalore, and the MPG-IISc Partner Group**. **H.K.** is grateful for the **PMRF fellowship**.

**H.K.** and **S.K.M.** express heartfelt gratitude to **Professor Rajeev Ranjan** for his generous guidance and unwavering support in addressing the reviewers' comments on this manuscript. His thoughtful technical insights and careful suggestions greatly enhanced the clarity and overall quality of the work, leading to a more comprehensive manuscript. **H.K.** and **S.K.M.** are also thankful to **Professor Dipankar Banerjee** and **Professor Abhik Choudhury** for fruitful technical discussions. **H.K.** is thankful to Ayush Kumar for Gibbs interfacial calculations, Dr. Chandan Kumar and Shashwat Kumar Mishra for their help in conducting mechanical tests.

**Contributions**

**S.K.M.** conceptualised the project and designed the experiments. **H.K.** prepared the alloys, carried out the microstructural and mechanical characterisation, and processed the data. **H.K.** and **S.K.M.** performed the analysis, interpreted the data, and wrote the manuscript. **P.K.** contributed to the interpretation of the mechanical deformation data. **B.G.** and **D.R.** contributed to the discussion and editing of the manuscript. All authors contributed extensively to the work and approved the final version of the manuscript.

# Supplementary Materials for

**High-strength and ductile lightweight cast aluminium alloys with superlattice nano-layered fibres (SNL) and core-shell nano-particles**

Hemant Kumar[1*], Praveen Kumar[1], Dierk Raabe[2], Baptiste Gault[2,3*], Surendra Kumar Makineni[1*]

[1]Department of Materials Engineering, Indian Institute of Science, Bangalore 560012

[2] Max Planck Institute for Sustainable Materials, 40237 Düsseldorf, Germany

[3]Department of Materials, Royal School of Mines, Imperial College, Prince Consort Road, London SW7 2BP, United Kingdom

***Corresponding authors:**

hemantkumar2@iisc.ac.in, b.gault@mpie.de, skmakineni@iisc.ac.in

**Fig. S1: Microstructure of As-cast Al-Gd (AG) alloy.** A Back-scattered-electron (BSE) image of the cast AG alloy showing the distribution of α-Al/$Al_3Gd$ eutectic structure and α-Al matrix.

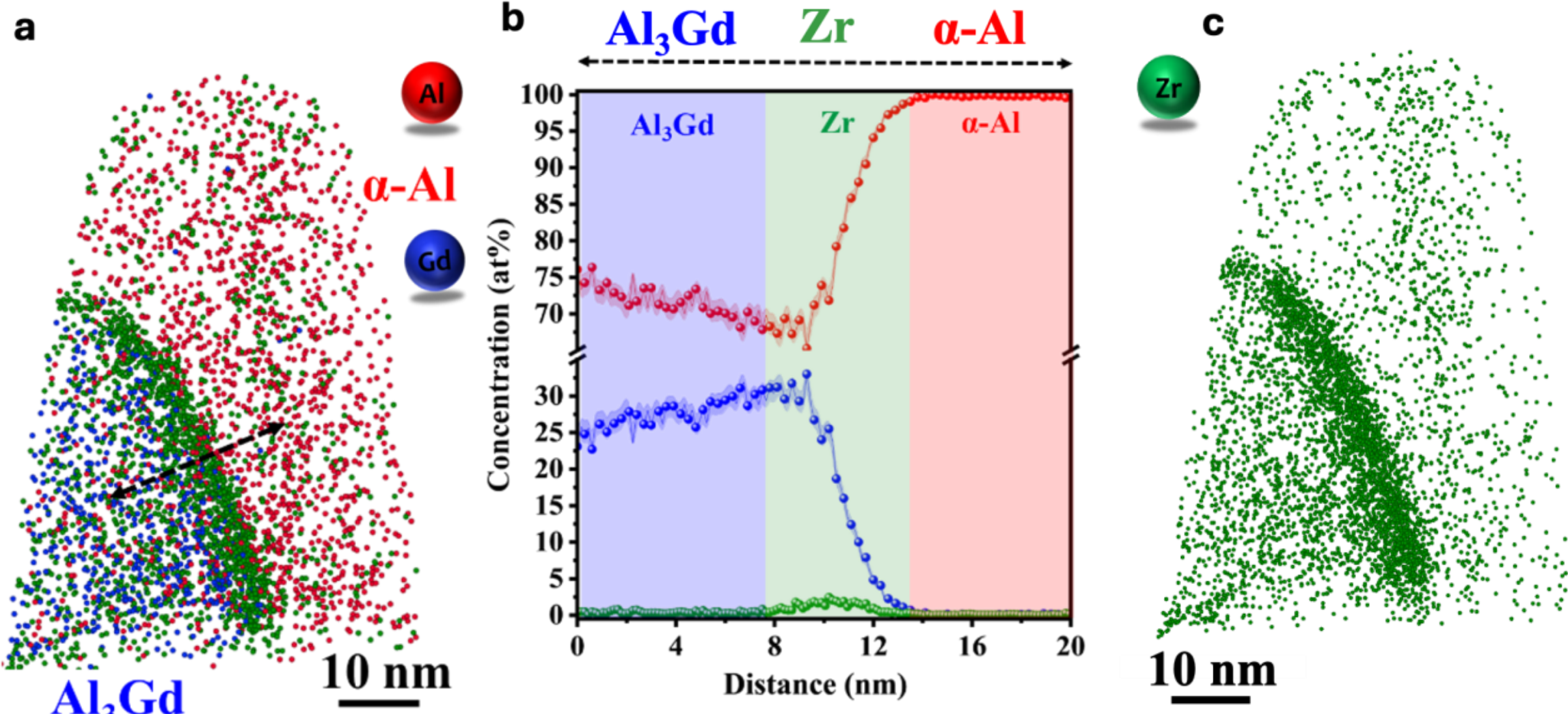

**Fig. S2: Atomic-scale elemental distribution across the α-Al/$Al_3Gd$ interface after 5 hours of annealing at 400°C in AGZ alloy: a** An atom probe reconstruction centered on an interface showing the distribution of Al (red atom), Gd (blue atom), and Zr (green atom). **b** The corresponding compositional profiles across the interface and **c** Zr distribution.

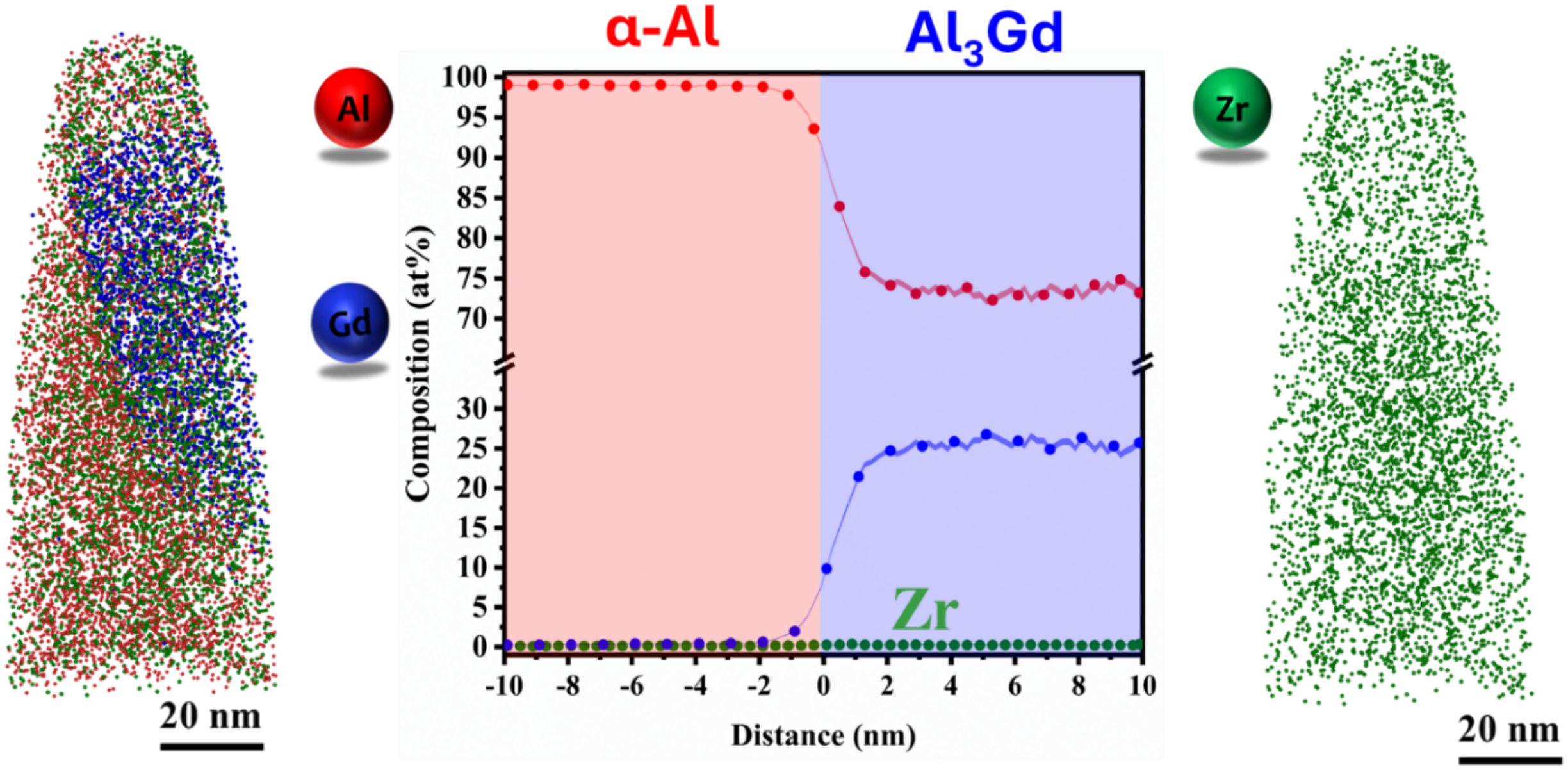

**Fig. S3: Atomic-scale elemental distribution across the α-Al/$Al_3Gd$ interface in as-cast AGZ alloy:** Atom probe reconstruction showing the distribution of Al (red atom), Gd (blue atom), and Zr (green atom) across the α-Al/$Al_3Gd$ fibre interface with the corresponding compositional profiles that reveal no Zr segregation at the interface.

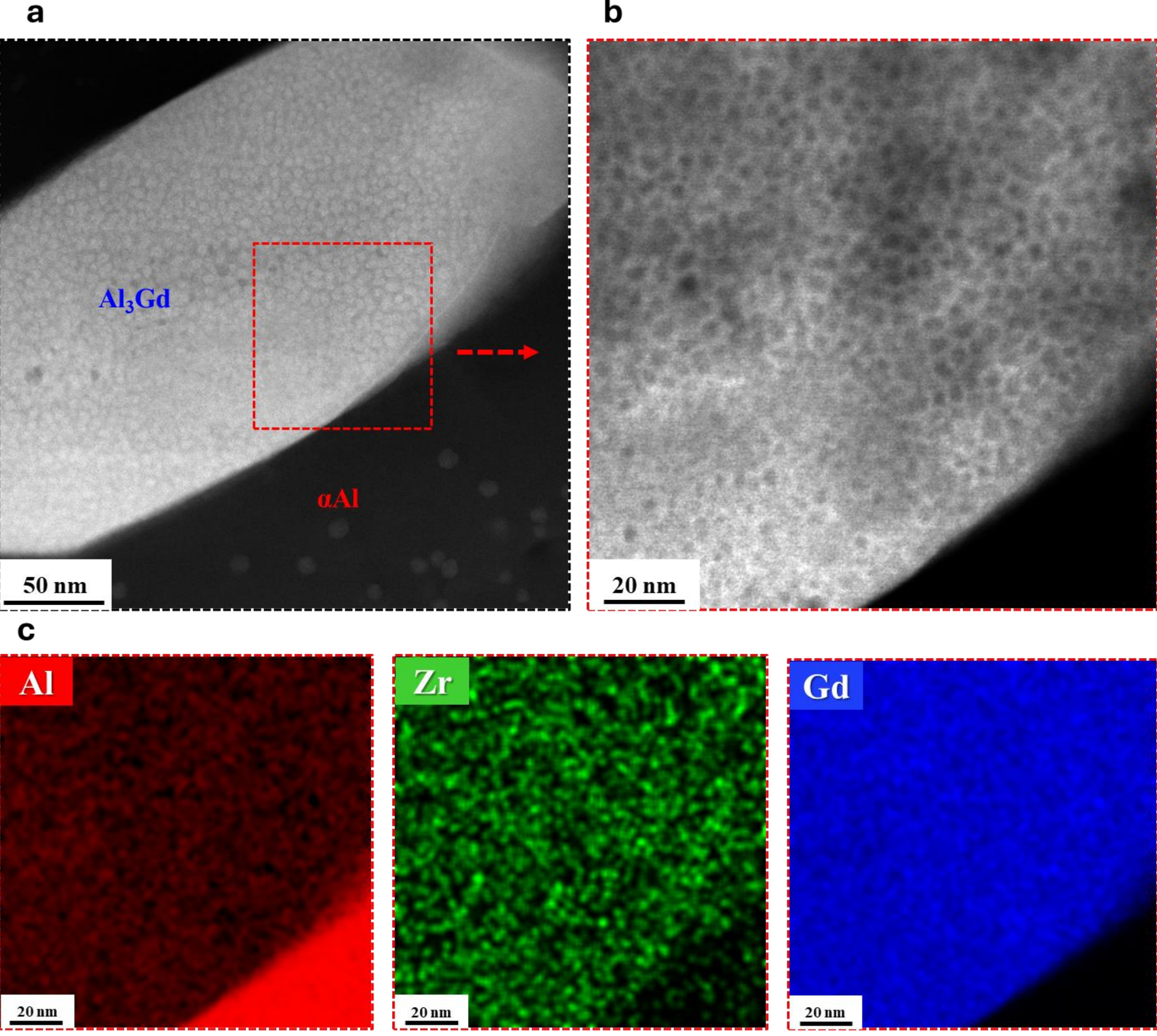

**Fig. S4: Top view of the SNL fibres in the Al-Gd-Zr alloy after annealing at 400°C for 25 hours: a** A low magnification high-angle-annular-darkfield (HAADF) STEM image centred on an SNL $Al_3Gd$ fibre. **b** Magnified image taken from the red dotted box showing the near conformal distribution of SNL on the $Al_3Gd$ fibre. **c** The energy-dispersive spectroscopy (EDS) elemental mapping across the region shows the distribution of Gd, Zr, and Al atoms, respectively.

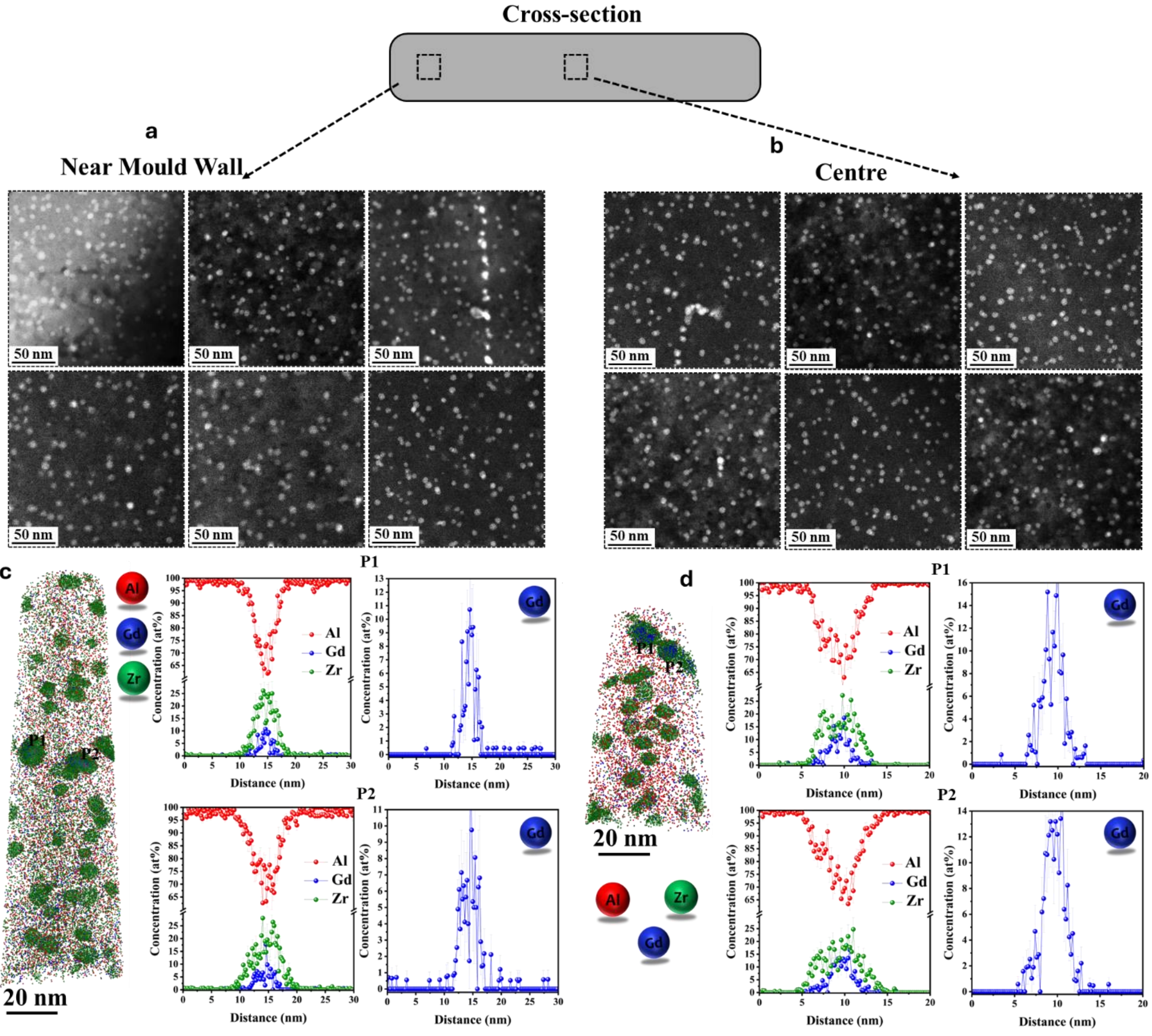

**Fig. S5:** High magnification HAADF STEM images showing the distribution of nano-particles in primary α-Al matrix from two locations: **a** near the mould wall and **b** at the Centre. Atom probe reconstructions from the primary α-Al matrix at the respective locations, showing the distribution of nano-particles and compositional profiles across core-shell nano-particles, showing an increase in Gd at the core up to ~12at.%.

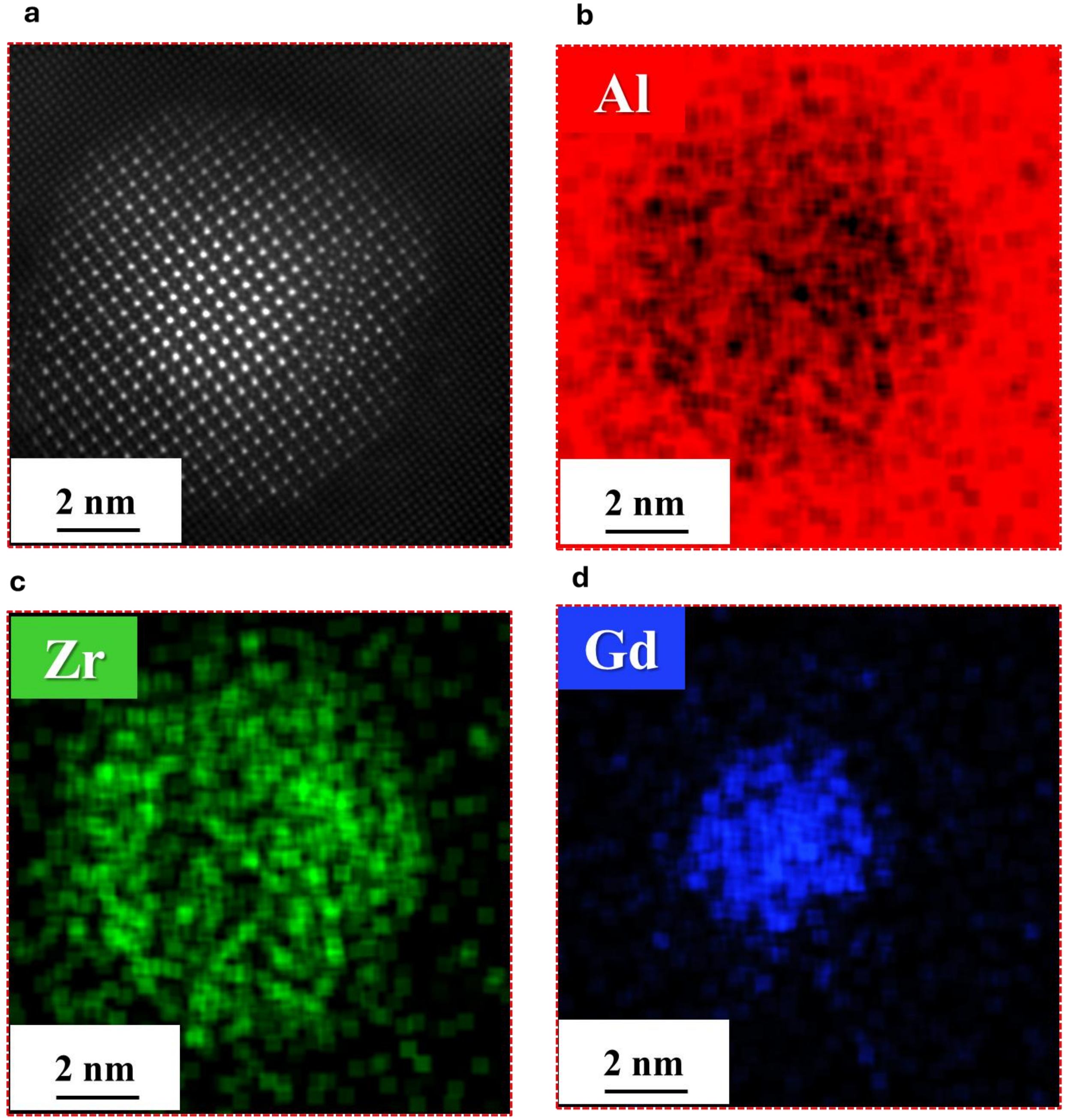

**Fig. S6: Energy Dispersive Spectroscopy (EDS) elemental map across a core-shell nano-particle: a** An HAADF-STEM image along [0 01] cubic zone axis centred on a core-shell nano-particle in α-Al matrix in the AGZ-SNL alloy, which reveals the relatively brighter contrast at the core of the nano-particle. This was further confirmed by the corresponding energy-dispersive spectroscopy (EDS) elemental mapping **b-d** across the core-shell nano-particle, which shows Gd enrichment at the core, while Zr is homogeneously distributed.

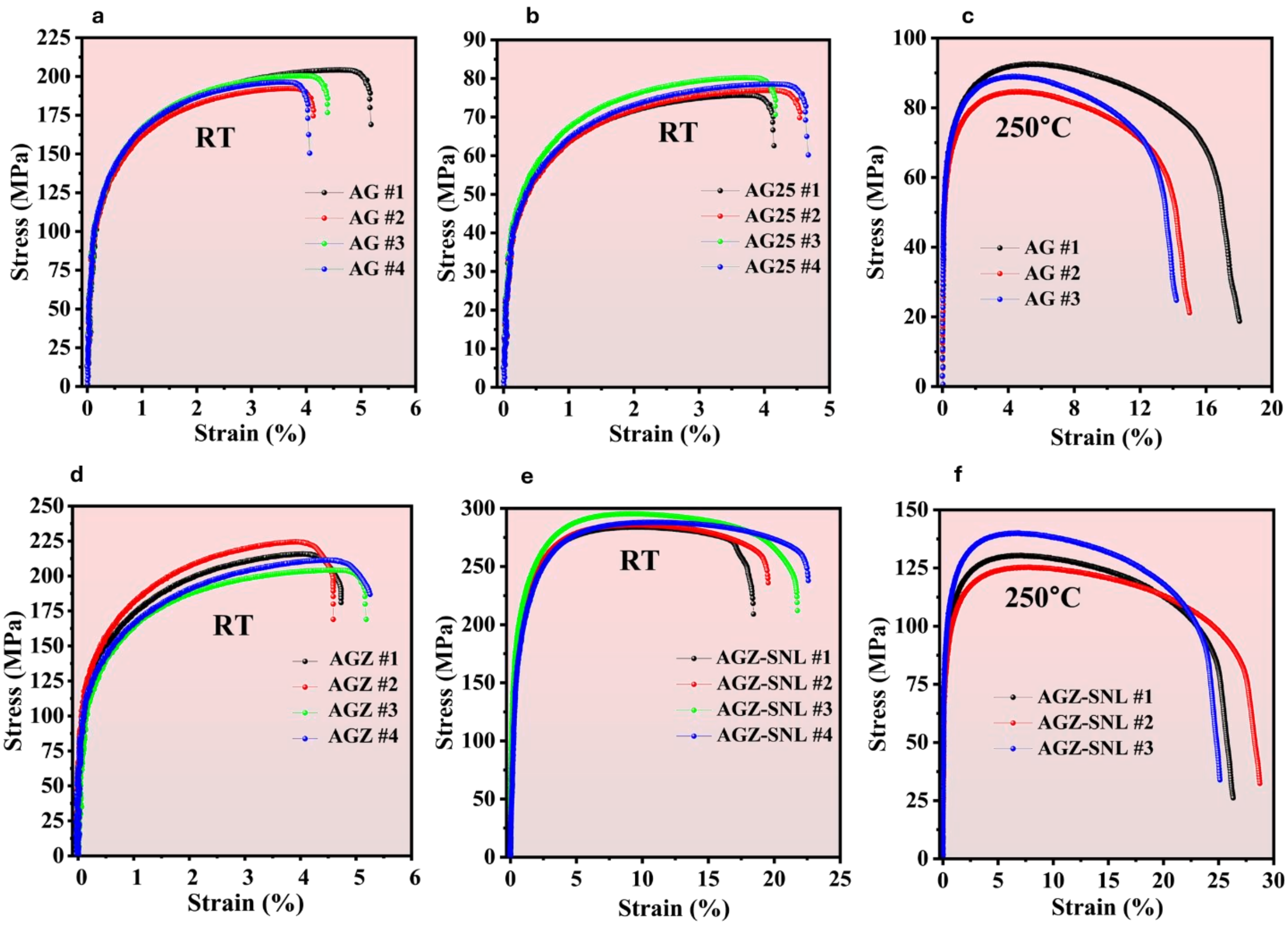

**Fig. S7: Engineering stress-strain plot for the all tested samples: a** AG alloy at RT, **b** AG alloy annealed at 400 °C for 25hr (AG25) at RT , **c** AG alloy at 250°C , **d** AGZ alloy at RT, **e** AGZ-SNL alloy at RT, and **e** AGZ-SNL alloy at 250°C.

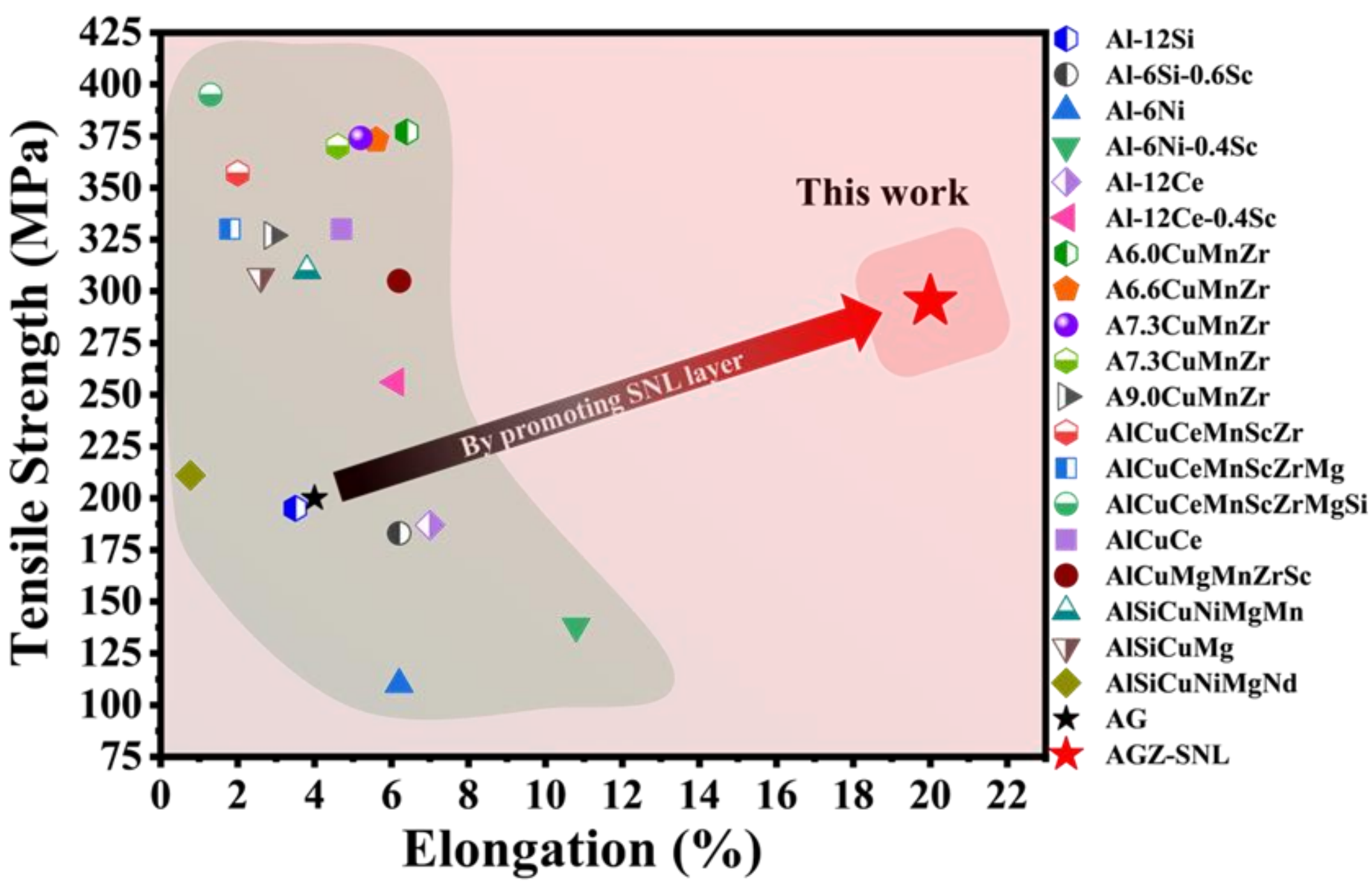

**Fig S8:** An Ashby plot between tensile strength and elongation for available high-strength cast Al eutectic alloys (Al-11.5Si[1], Al-5.8Si-0.36Sc[2], Al-2.9Ni[3], Al-2.9Ni-0.25Sc[3], Al-2.7Ce[4], Al-2.7Ce-0.3Sc[4], Al-Cu-Mn-based[5], Al-Cu-Ce-based[6,7], Al-Cu-Mg-based[8], and Al-Si-Mg-Cu-Ni-based[28-30]) in comparison to Al-2.5Gd (AG) and Al-2.5Gd-0.15Zr (AGZ-SNL) alloys.

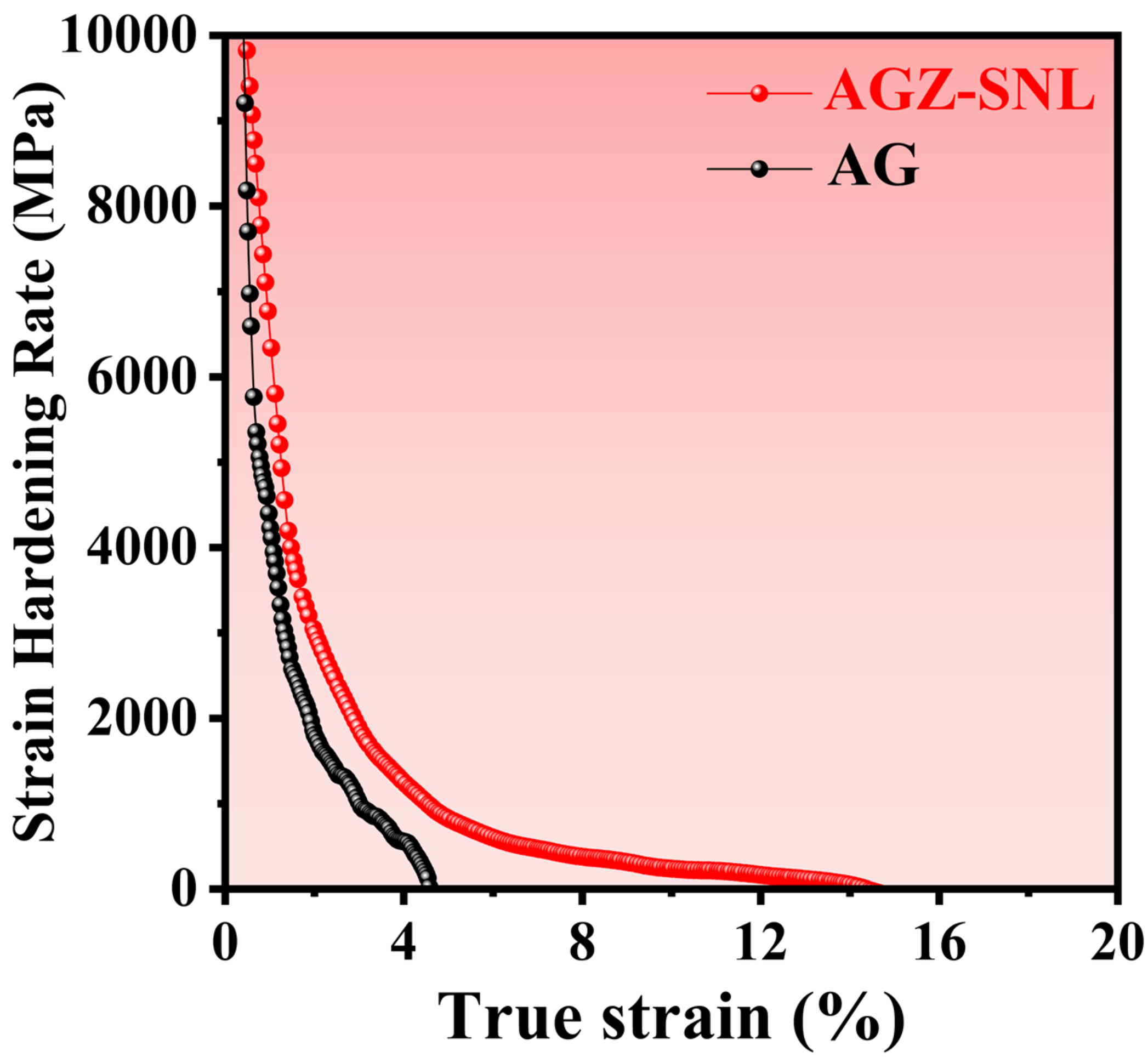

**Fig. S9:** Comparison of Strain Hardening rate (MPa) vs True Strain (%) plots for AG and AGZ-SNL alloys.

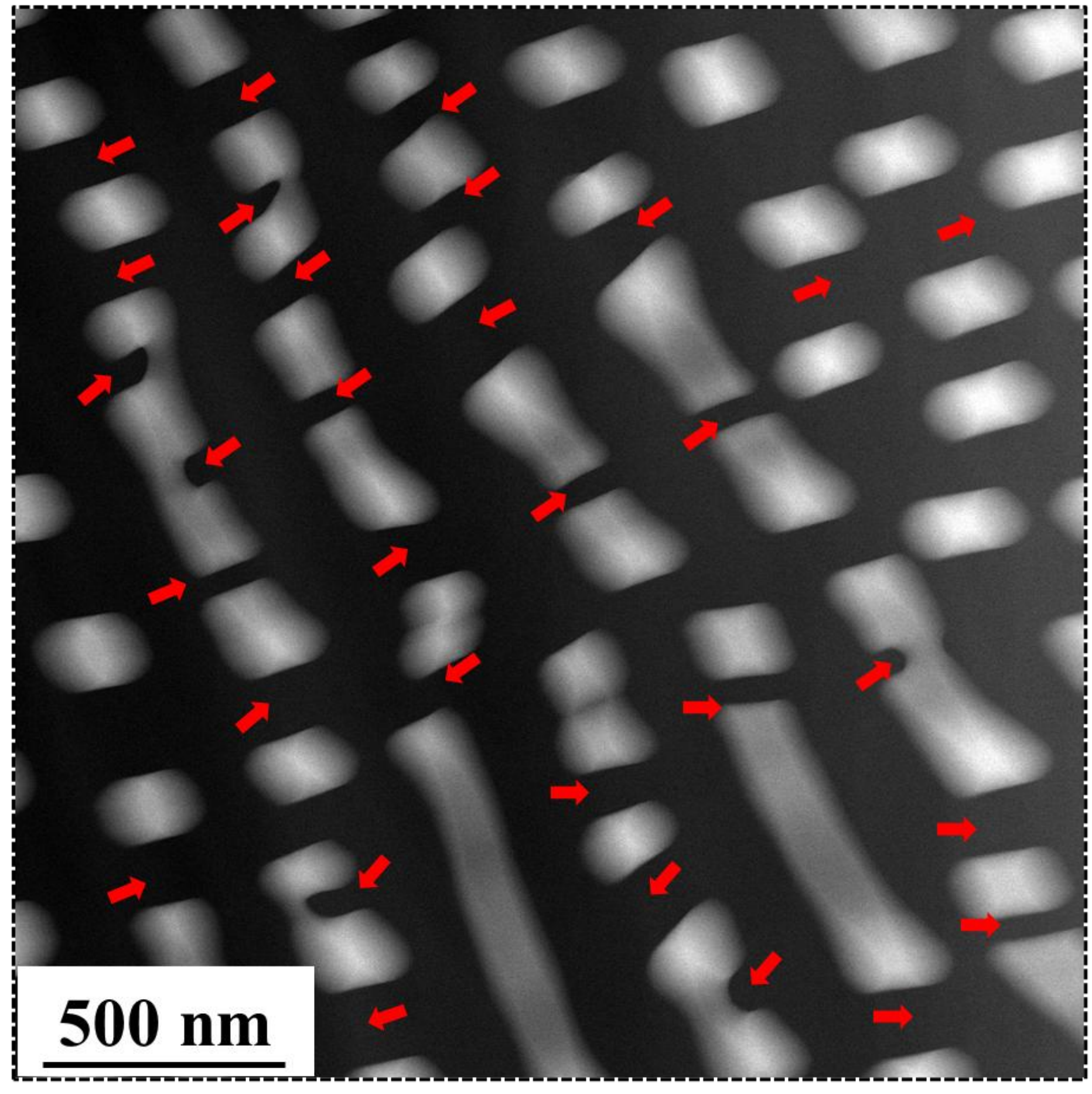

**Fig. S10:** Microstructure of eutectic region after fracture of AG alloy showing multiple cracks that are propagating across the brittle fibres.

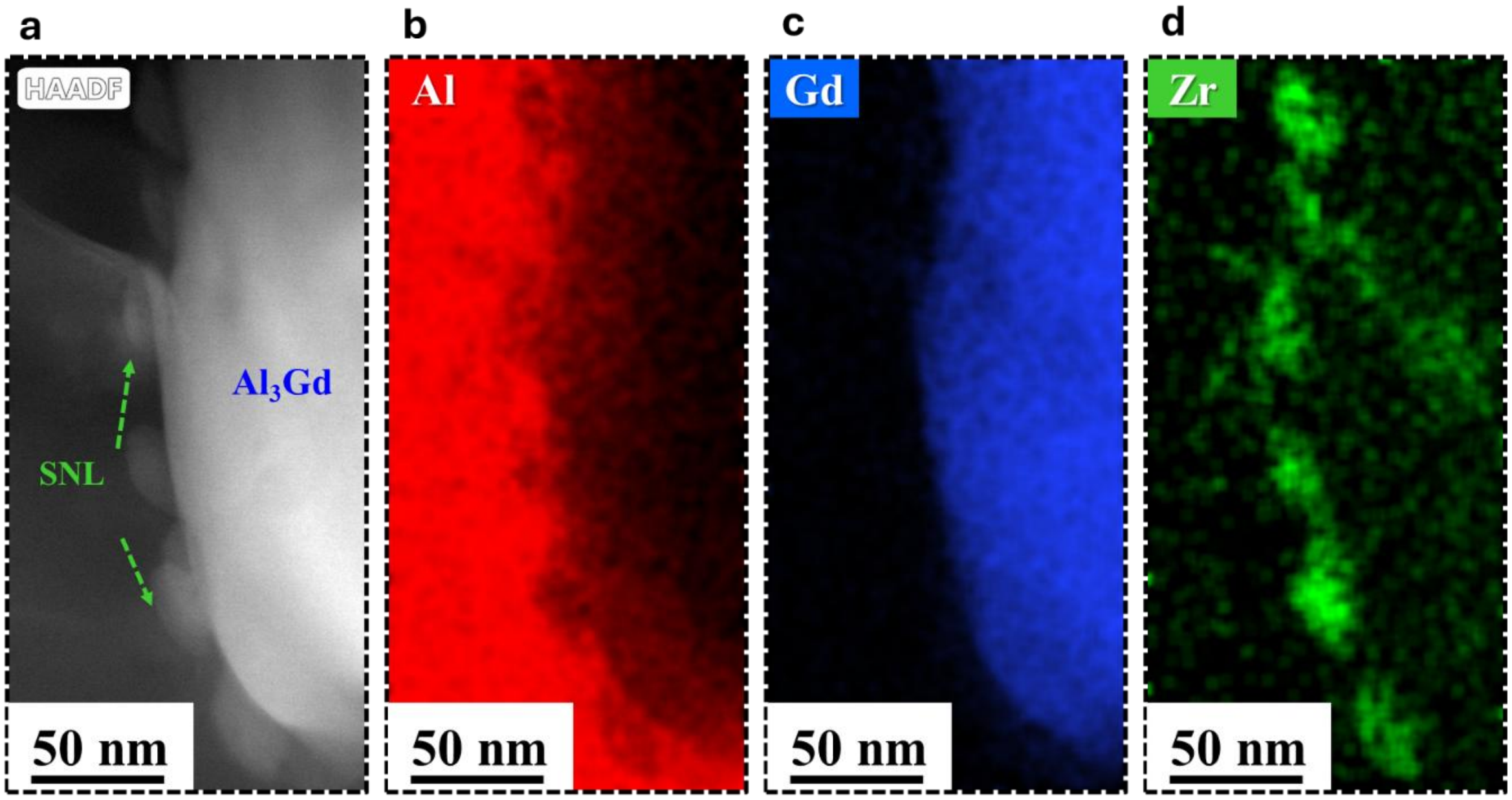

**Fig. S11: EDS elemental map across the interface of SNL fibre after tensile fracture of AGZ-SNL alloy at 250°C. a** A low magnification HAADF-STEM image, from the interface of SNL fibre, showing the corresponding **b-d** EDS elemental mapping of Al, Zr, and Gd.

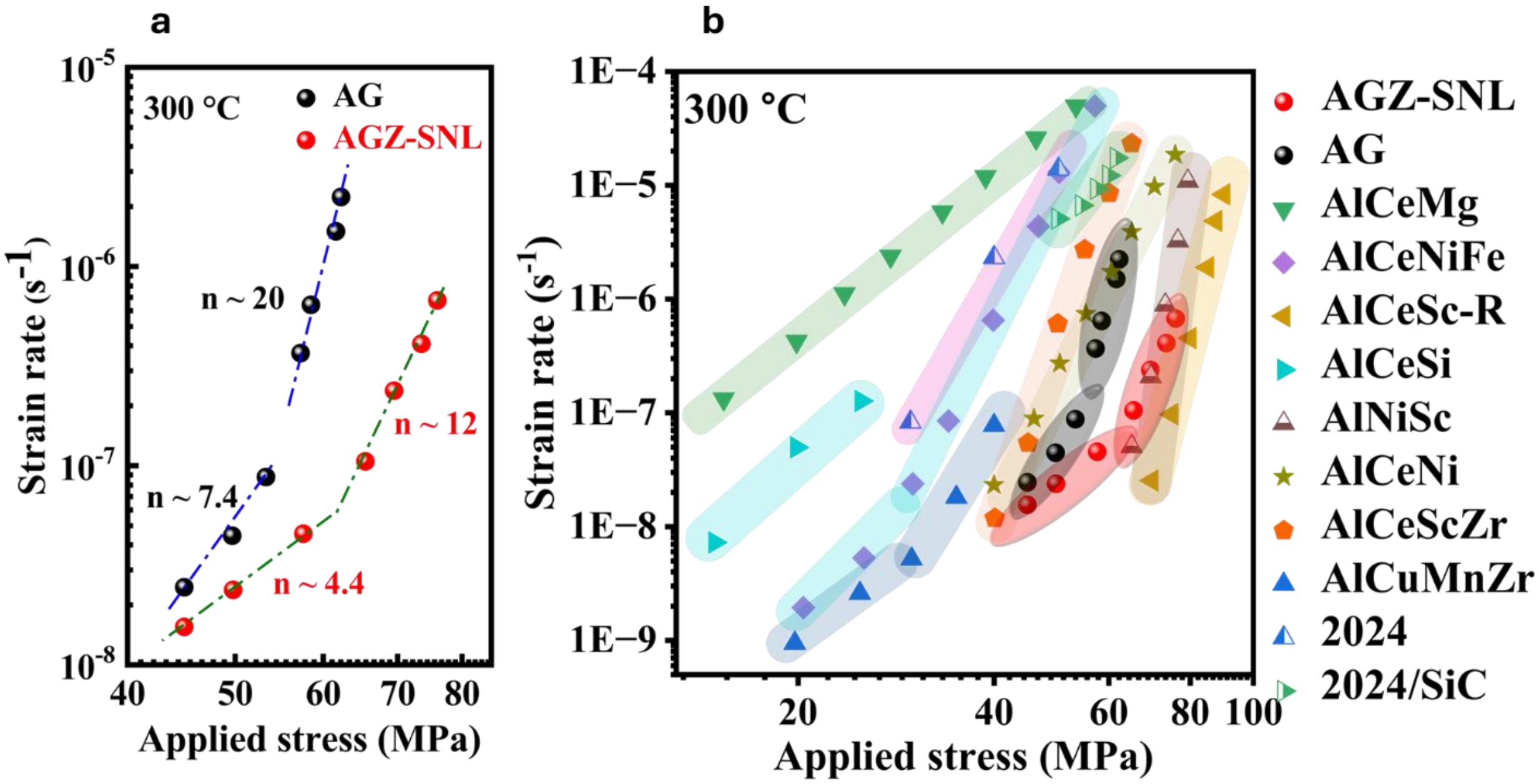

**Fig. S12: Comparison of creep properties of AG, AGZ-SNL, and other Al eutectic and precipitation hardenable alloys. a** A plot of steady state compressive creep strain rate ($s^{-1}$) as a function of applied stress tested at constant temperature (300°C) of AGZ alloy annealed at 400°C for 25 hours in comparison with AG cast alloy. **b** A plot of steady state compressive creep strain rate ($s^{-1}$) as a function of applied stress of AGZ alloy tested at 300°C in comparison with previously reported Al-Ce based eutectic alloys (AlCeMg[1], AlCeNiFe[2], AlCeSc[3], AlCeSi[4], AlCeNi[5], AlCeScZr[6], Al-Ni-Sc[7], Al-Cu-Mn-Zr[8], 2024[9] and 2024/SiC[10] precipitation-hardenable alloys. The AGZ alloy with SNL and core-shell nano-particles exhibits a nearly two orders of magnitude lower creep rate than the AG alloy at an applied stress of ~ 60 MPa. Compared to other alloys, the creep rate is several orders of magnitude lower for AGZ-SNL alloy, except nearly similar to Al-Ce-Sc and Al-Ni-Sc.

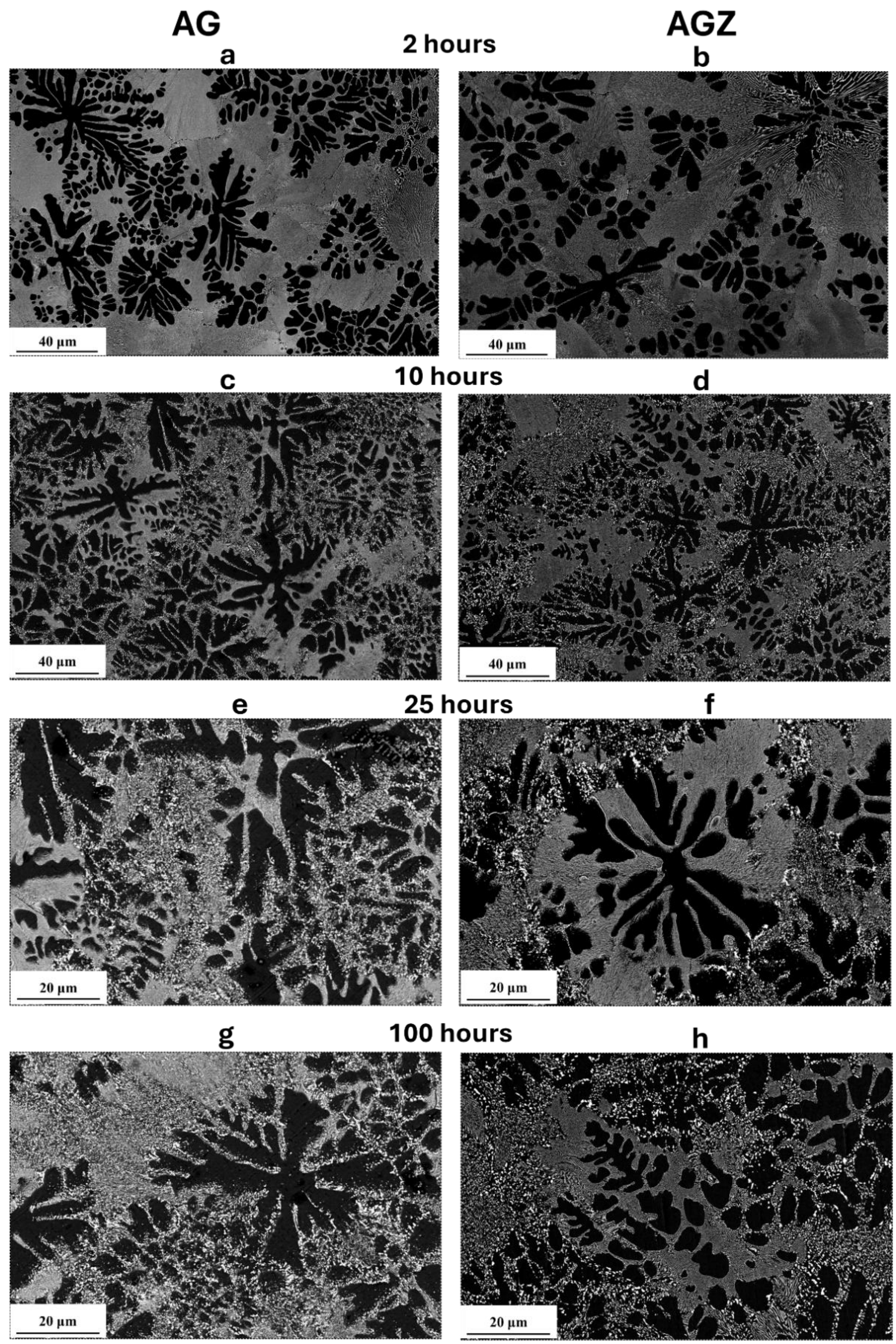

**Fig. S13:** Comparison of microstructures of AG and AGZ alloys annealed at 400°C for **a,b** 2 hours, **c,d** 10 hours, **e,f** 25 hours and **g,h** 100 hours. (Brighter contrast regions in the images are the coarse eutectic microstructure)

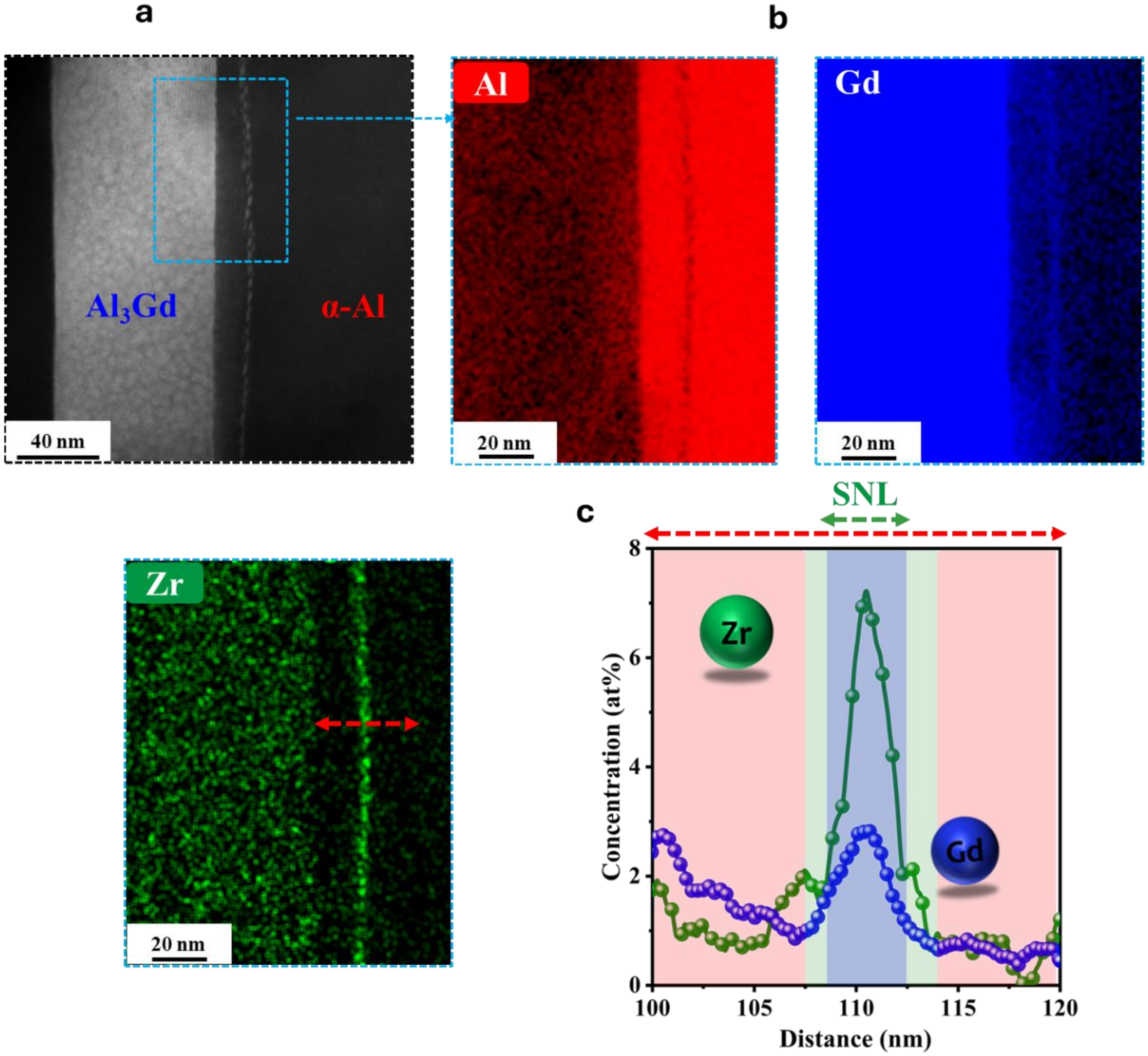

**Fig. S14: EDS elemental map across the Disordered Zone (DZ) and SNL of AGZ-SNL alloy. A HAADF STEM image centered on an Al3Gd interface showing the presence of a** disordered zone (DZ) between the fiber and SNL. **b** The corresponding EDS elemental map and **c** profiles of Zr and Gd across the DZ, which shows the SNL has a Zr:Gd ratio of 3:1

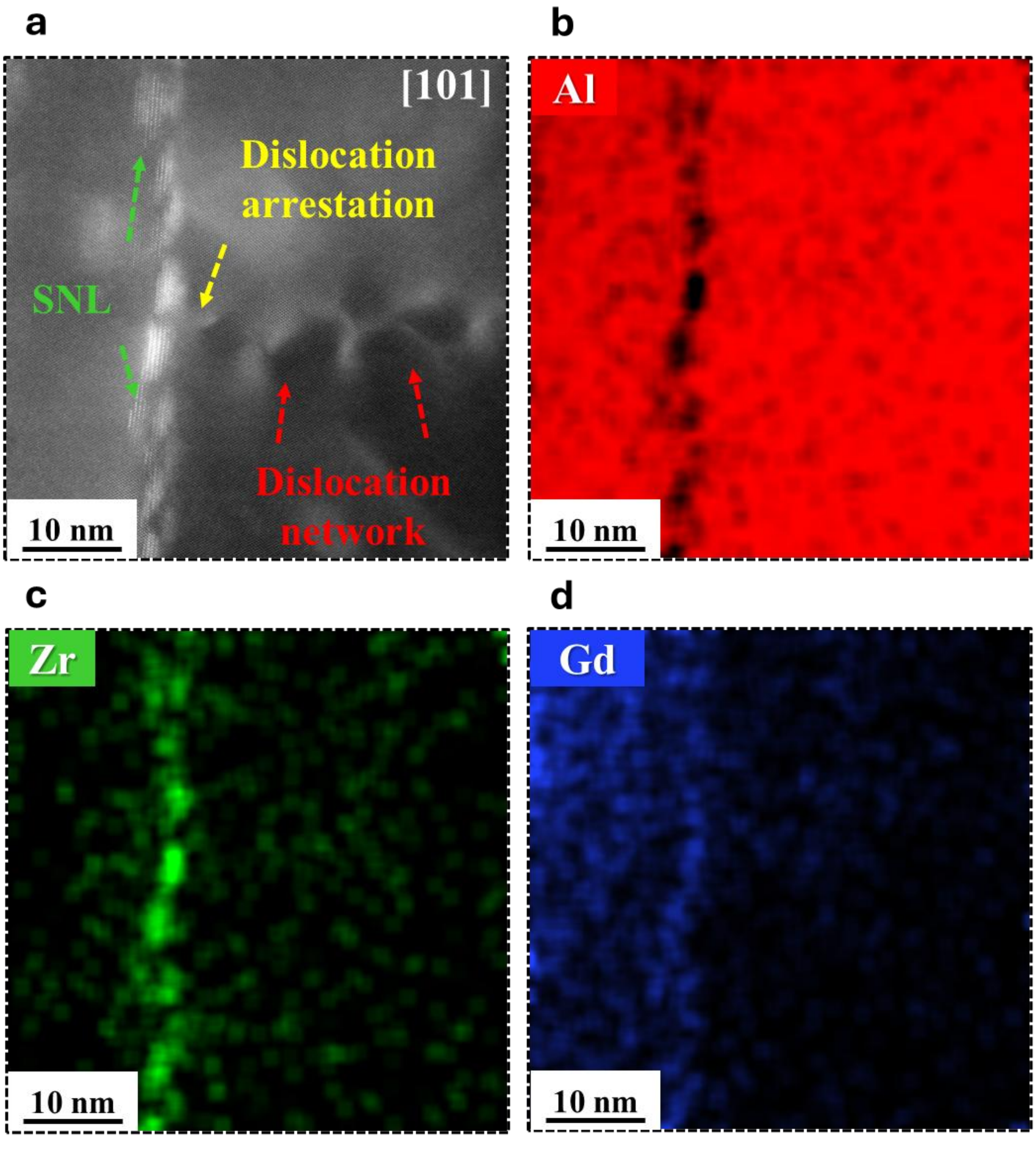

**Fig. S15:** EDS elemental map across a few accommodated dislocations near the SNL in a deformed Al-Gd-Zr alloy.

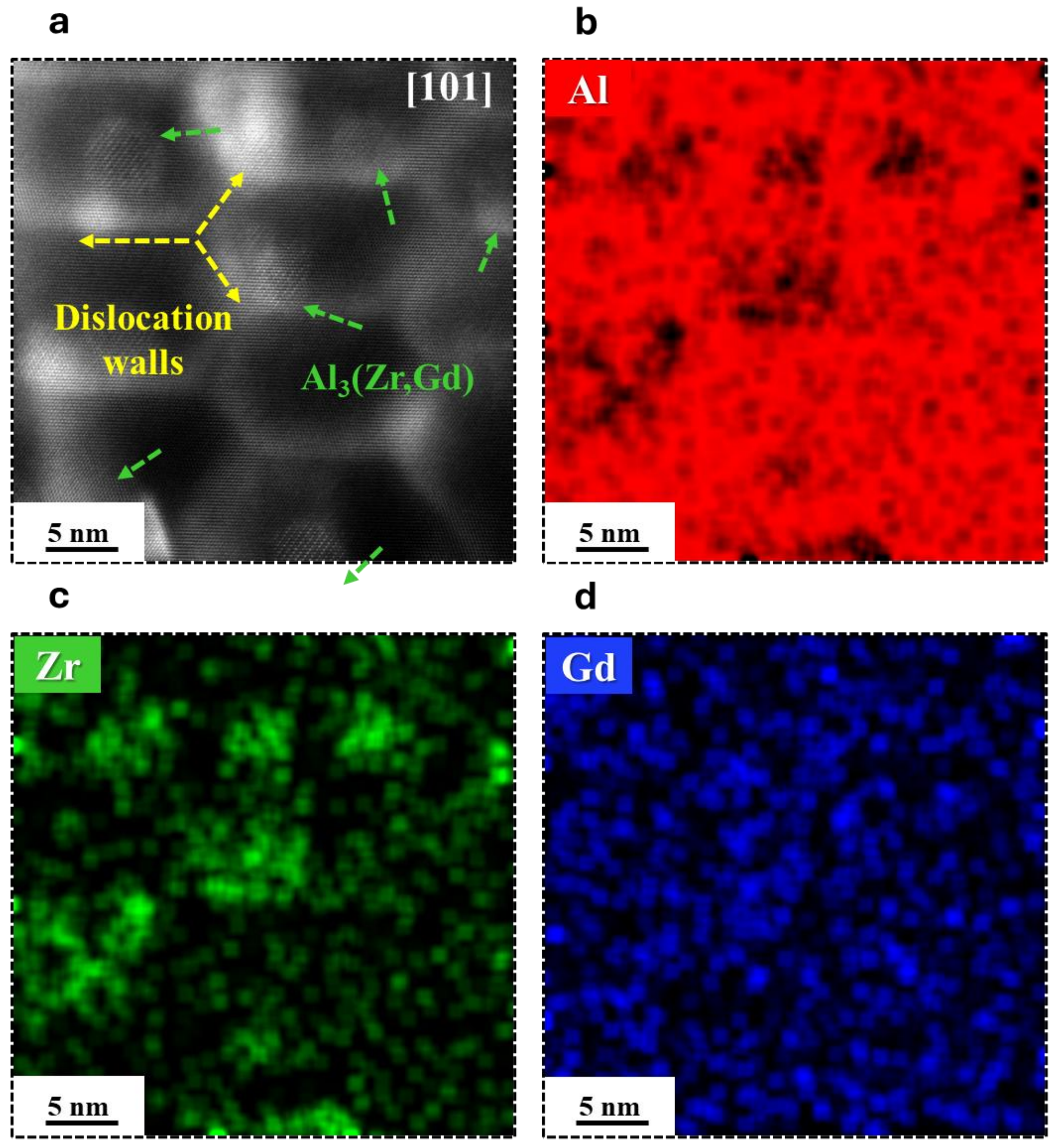

**Fig. S16:** EDS elemental map across the dislocation network in the primary α-Al matrix of the Al-Gd-Zr alloy after deformation

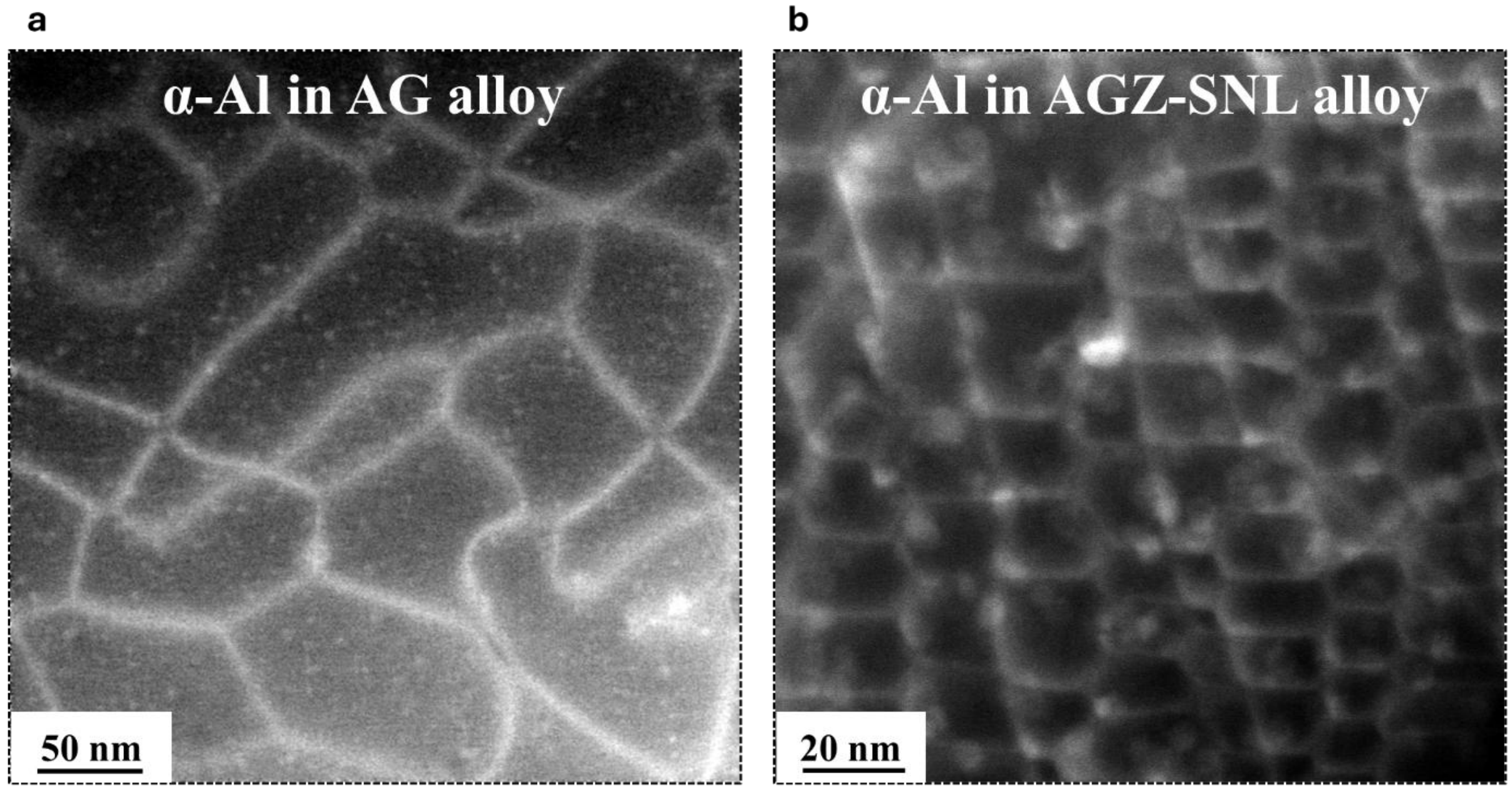

**Fig. S17:** LAADF images from the primary α-Al regions of AG and AGZ-SNL alloy after deformation showing relatively dense distribution of dislocation network cells in the latter**.**

| Alloys Nomenclature | Yield Strength, YS (MPa) | Ultimate Tensile Strength, UTS (MPa) | Elongation, EL (%) |
|---|---|---|---|
| AG | $125 \pm 2$ | $199 \pm 4$ | $4.5 \pm 0.6$ |
| AG 25 | $49 \pm 1.5$ | $78 \pm 2$ | $4.3 \pm 0.2$ |
| AGZ | $135 \pm 8$ | $213 \pm 8$ | $5 \pm 0.30$ |
| AGZ SNL | $190 \pm 10$ | $295 \pm 4$ | $21 \pm 1.5$ |
| AG-250°C | $63 \pm 2$ | $89 \pm 3$ | $15 \pm 1.5$ |
| AGZ SNL-250°C | $88 \pm 6$ | $131 \pm 6$ | $25 \pm 1.5$ |

**Fig. S18:** Tensile properties of all the alloys

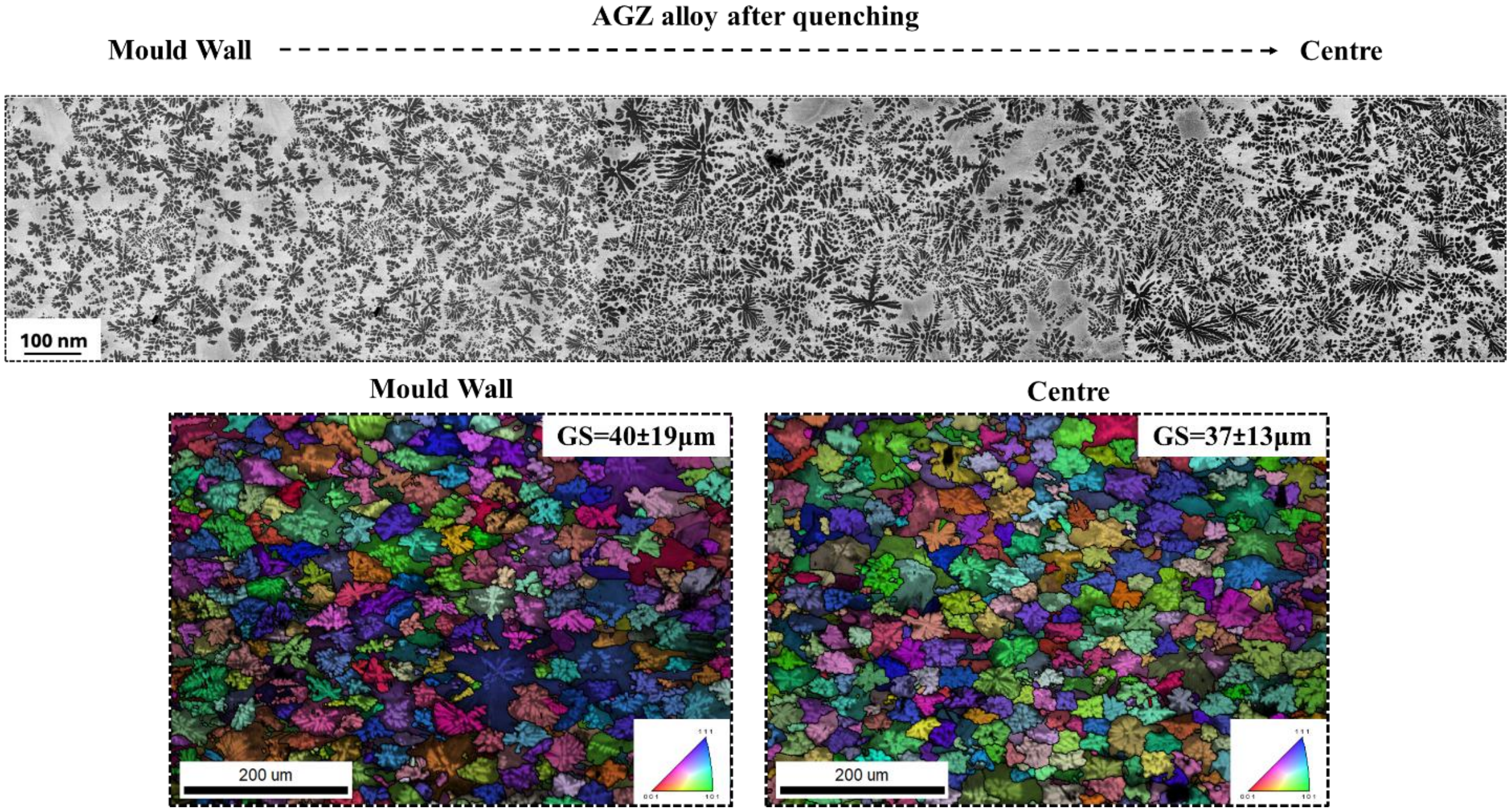

**Fig. S19:** Collage of images the collage of images taken near the mould wall to the centre of the cast heat-treated slab after quenching. EBSD IPF+IQ maps showing the grain structure from the region near the mould wall and centre.

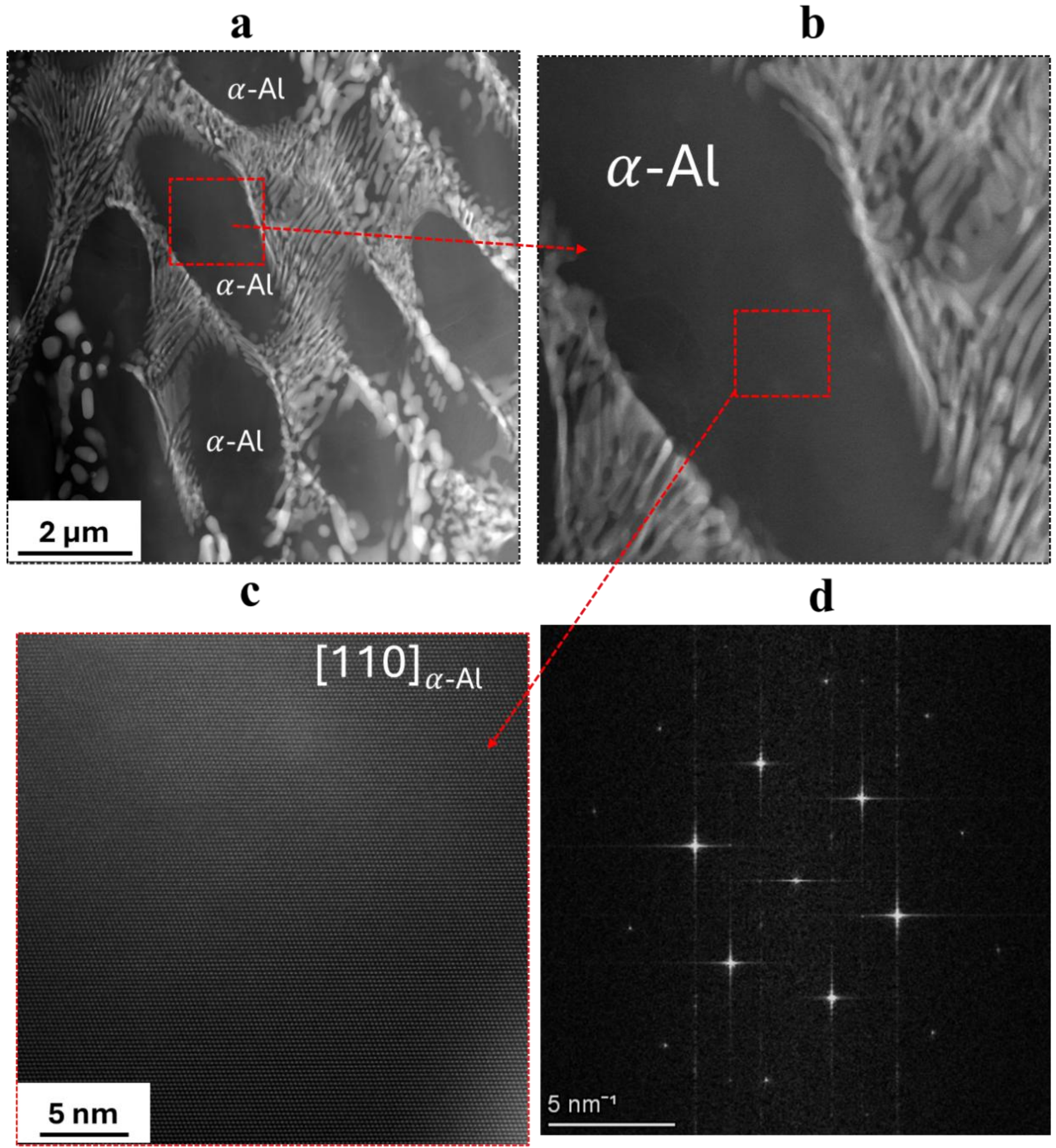

**Fig. S20: a.** A low magnification HAADF STEM image of cast Al-Gd-Zr alloy showing the distribution of primary α-Al and eutectic structure (α-Al/$Al_3Gd$). **b** A high magnification HAAD image centred on the primary α-Al showing the absence of nano-particles with ordered $L1_2$ structure inside the primary α-Al. **c** A high-resolution atomic-scale HAADF image from the same region and the corresponding **d**. diffraction pattern taken along [110] zone axis. The absence of any contrast and the superlattice spots in the diffraction pattern indicate that the nano-particles are not present in the primary α-Al region of the cast Al-Gd-Zr (AGZ) alloy.